\documentclass[12pt]{elsarticle}
\usepackage[latin9]{inputenc}
\usepackage{geometry}
\usepackage{amsmath}
\usepackage{amssymb}
\usepackage{graphicx}

\makeatletter

\usepackage{physics}
\usepackage{dcolumn}
\usepackage{bm}
\usepackage{xcolor}
\usepackage{array}
\usepackage{float}

\ifdefined\showcaptionsetup
 \PassOptionsToPackage{caption=false}{subfig}
\fi
\usepackage{subfig}
\makeatother

\begin{document}
\begin{frontmatter}
\title{A robust fourth-order finite-difference discretization for the strongly
anisotropic transport equation in magnetized plasmas}
\author[lanl]{L. Chac\'on\corref{cor1}}
\ead{chacon@lanl.gov}
\author[lanl]{Jason Hamilton}
\author[lanl]{Natalia Krasheninnikova}
\cortext[cor1]{Corresponding author}
\address[lanl]{Los Alamos National Laboratory, Los Alamos, NM 87545, USA}
\begin{abstract}
We propose a second-order temporally implicit, fourth-order-accurate
spatial discretization scheme for the strongly anisotropic heat transport
equation characteristic of hot, fusion-grade plasmas. Following {[}Du
Toit et al., \emph{Comp. Phys. Comm}., \textbf{228} (2018){]}, the
scheme transforms mixed-derivative diffusion fluxes (which are responsible
for the lack of a discrete maximum principle) into nonlinear advective
fluxes, amenable to nonlinear-solver-friendly monotonicity-preserving
limiters. The scheme enables accurate multi-dimensional heat transport
simulations with up to seven orders of magnitude of heat-transport-coefficient
anisotropies with low cross-field numerical error pollution and excellent
algorithmic performance, with the number of linear iterations scaling very weakly with grid resolution and grid anisotropy, and scaling with the square-root of
the implicit timestep. We propose a multigrid preconditioning strategy
based on a second-order-accurate approximation that renders the scheme
efficient and scalable under grid refinement. Several numerical tests
are presented that display the expected spatial convergence rates
and strong algorithmic performance, including fully nonlinear magnetohydrodynamics
simulations of kink instabilities in a Bennett pinch in 2D helical
geometry and of ITER in 3D toroidal geometry.
\end{abstract}
\end{frontmatter}

\section{Introduction}

This study is concerned with the solution of the strongly anisotropic
transport equation in magnetized plasmas. Plasmas in tokamak (and
similar fusion devices) rely on a dominant axial field that limits
cross-field transport, which would otherwise degrade the confinement
by allowing energy to escape to the walls. Theoretical estimates,
experimental measurements, and modeling suggest that the transport
anisotropy in common tokamak reactors can reach extremely high values
$\chi_{\parallel}/\chi_{\perp}\sim10^{7}$ --- $10^{10}$ \citep{braginskii,holzl2009determination,ren1998measuring,meskat2001analysis,snape2012influence,choi2014improved},
making solving this equation a formidable numerical challenge. Such
strong magnetization leads to much higher transport anisotropy than
that found in other applications, and thus presents unique numerical
challenges for simulating realistic plasma conditions in simulations.

The numerical challenges are present both spatially and temporally.
Spatially, strongly anisotropic diffusion equations suffer from numerical
pollution of the perpendicular dynamics from the large parallel transport
term \citep{umansky2005numerical}, and the lack of a maximum principle
(which may lead to negative temperatures). Numerical pollution can
be mitigated by the use of high-order spatial discretizations \citep{sovinec2004nonlinear,gunter2007finite}
and/or so-called ``symmetric'' discretizations \citep{gunter2005modelling,gunter2007finite,van2014finite,van2016finite},
but these lack a maximum principle. The maximum principle can be enforced
via nonlinear limiters \citep{sharma-jcp-07-anis,lepotier2008finite,kuzmin-jcp-09-anis,lipnikov2012minimal,dutoit2018positivity}
or via locally identifying diffusion directions \citep{thangaraj1998rotated,lepotier2009schema,ngo2016monotone},
but the former typically result in low-order spatial discretizations
(and, depending on the limiter used, strong nonlinearity), and the
latter lose local conservation, of critical importance in many applications
of interest. Temporally, explicit methods are constrained to very
small time steps due to the Courant stability condition, determined
by the largest diffusion coefficient (i.e., the parallel transport
one). For implicit methods, the issue is the near-degeneracy of the
associated algebraic systems due to the strong transport anisotropy
\citep{chacon2014asymptotic}, and the ill-conditioning of associated
algebraic systems that makes them difficult to invert using modern,
efficient iterative methods. Most of the previous studies using implicit
methods resorted to direct solvers for the temporal update, with some
authors resorting to implicit/explicit integrators to ameliorate the
size of the resulting matrices on a per-parallel-task basis \citep{gunter2009mixed}.
More recent studies \citep{li2024block} have explored approximate-block-factorization
preconditioners for GMRES for the micro-macro formulation of Ref.
\citep{degond2012asymptotic}, demonstrating significant potential
for controlling the number of GMRES iterations but so far only for
mesh-aligned and simply connected magnetic-field topologies (i.e,
without islands).

Recently, a series of studies have proposed asymptotic-preserving
(AP) schemes for the anisotropic transport equation \citep{degond2010duality,degond2010asymptotic,degond2012asymptotic,narski2014asymptotic,wang2018uniformly}
with the property that the numerical error and matrix conditioning
do not scale with the anisotropy ratio $\chi_{\parallel}/\chi_{\perp}$.
Refs. \citep{degond2010duality,degond2010asymptotic,degond2012asymptotic}
considered only open field lines in a time-independent context. In
contrast, Ref. \citep{wang2018uniformly} considered only closed ones,
also in a time-independent context. Ref. \citep{narski2014asymptotic}
considered the time-dependent case for open and closed magnetic fields
with implicit timestepping. However, it is unclear how the approach
can generalize to three dimensions (where confined stochastic field
lines of infinite length may exist), and the reference employed a
direct linear solver, which is known to scale very poorly with mesh
refinement and with processor count in parallel environments.

A separate AP line of research is a semi-Lagrangian scheme based on
a Green's function formulation of the heat transport equation, where
parallel fast transport is resolved essentially semi-analytically
\citep{del2011local,del2012parallel,chacon2014asymptotic,chacon-jcp-24-arB_xport,chacon-jpc-24-imp_xport}.
Those methods ensure the absence of pollution due to their asymptotic
preserving nature \citep{chacon2014asymptotic} in the limit of infinite
anisotropy \citep{del2011local,del2012parallel}, can deal with nonlocal
heat closures \citep{chacon2014asymptotic} and arbitrary magnetic
field topology \citep{chacon-jcp-24-arB_xport}, and allow for implicit
timestepping \citep{chacon-jpc-24-imp_xport}. However, to date, these
methods need to be demonstrated with nonlinear transport coefficients, have not
been extended to deal with general boundary conditions for the magnetic
field, and have not yet been coupled with richer physics models such as
magnetohydrodynamics (MHD).

This study proposes a practical implicit Eulerian spatio-temporal
discretization for the strongly anisotropic transport equation which,
while not AP, features several desirable properties, including:
\begin{itemize}
\item Manageable numerical pollution for reasonable anisotropies, demonstrated
here up to $\chi_{\parallel}/\chi_{\perp}\sim10^{7}$  (which
in 2D correspond to $\sim10^{9}$ anisotropies in 3D for large-magnetic-guide
field configurations such as tokamaks \citep{gunter2009mixed}).
\item Numerical robustness against developing negative temperatures for
fourth-order accurate discretizations, and strict positivity for second-order
ones.
\item Strict local conservation properties.
\item Suitability for modern nonlinear solvers (e.g., Jacobian-free Newton-Krylov
-- JFNK \citep{knoll-jcp-04-nk_rev}) with efficient multigrid preconditioning
for scalability.
\item Compatibility and ease of implementation in existing finite-difference
multiphysics simulation codes (e.g., MHD).
\end{itemize}
Our implementation employs nonlinear flux limiters following the key
insight proposed in Ref. \citep{dutoit2018positivity} (in the context
of the Fokker-Planck collision operator). The reference proposed to
reformulate mixed-derivative terms in the diffusion operator as nonlinear
advection operators, to which positivity-preserving limiters that
are compatible with nonlinear iterative solvers can be applied (here
we use a variant of the SMART advective scheme \citep{smart}). Our
contributions beyond that study include:
\begin{itemize}
\item Extension to fourth-order accuracy (critical for dealing with strong
anisotropies, as we shall see).
\item Development of effective and scalable multigrid preconditioning strategies
that render the linear and nonlinear iteration count manageable for
sufficiently large timesteps (measured as $\Delta t\chi_{\parallel}$).
\item Demonstration on challenging anisotropic transport problems, including
full 3D MHD simulations of fusion-grade plasmas.
\end{itemize}
The rest of this paper is organized as follows. The model of interest
along with spatial and temporal discretization details is introduced
in Sec. \ref{sec:equations}. Section \ref{sec:numtests} demonstrates
the properties of the scheme with several challenging numerical tests,
including fully featured MHD simulations of kink instabilities in
a Bennett pinch \citep{bennett1934magnetically} in 2D, and in the
ITER fusion reactor in 3D. We finally conclude with Sec. \ref{sec:Discussion-and-conclusions}.

\section{Model equations, numerical discretization, and solver strategy}

\label{sec:equations}The anisotropic heat transport equation in magnetized
plasmas reads: 
\begin{equation}
\partial_{t}T-\nabla\cdot(\bar{\bar{\Xi}}(T)\cdot\nabla T)\equiv S,\label{diffusion_eq}
\end{equation}
where $T=T(t,\mathbf{x})$ is the temperature field, $S=S(t,\mathbf{x})$
is a heat source, and the tensor diffusion $\bar{\bar{\Xi}}(T)$ is
given by:
\[
\bar{\bar{\Xi}}(T)=\chi_{\parallel}(T)\mathbf{b}\mathbf{b}+\chi_{\perp}(T)(\mathbb{I}-\mathbf{b}\mathbf{b}),
\]
with $\mathbf{b}=\mathbf{B}/B$ the unit vector along the magnetic
field direction. The expression for the parallel and perpendicular
transport coefficients for collisional fully ionized plasmas can be
found in Ref. \citep{braginskii}, and will be discussed later in
this study. The extreme transport anisotropy, $\chi_{\parallel}/\chi_{\perp}\sim10^{7}-10^{10}$,
makes finding the solution of this equation a formidable numerical
challenge (as advanced in the previous section).

In this study, we propose an accessible higher-order finite-difference
spatial discretization with remarkable robustness and solver performance
in an implicit timestepping context. Key to the approach is the reformulation
of mixed-derivative terms as nonlinear advection operators \citep{dutoit2018positivity}.
In what follows, we introduce the main ingredients of the nonlinear
spatial discretization and its nonlinear iterative inversion, including
its linearized treatment to make the approach amenable to multigrid
preconditioning.

\subsection{Fourth-order discretization of the anisotropic diffusion equation}

In this section we will develop a conservative, fourth-order discretization
framework for the $\nabla\cdot\left(\bar{\bar{\Xi}}\cdot\nabla T\right)$
operator. This discretization has been used in various earlier studies
\citep{chacon2014asymptotic,chacon-jpc-24-imp_xport,chacon-jcp-24-arB_xport},
but has never been documented in detail and we do so here. We will
derive the formulas in detail for the 2D Cartesian uniform-mesh case,
noting that the extension to 3D is straightforward. We also note that
the extension to curvilinear geometry is also straightforward when
realizing that, in curvilinear representation $\mathbf{x}(\boldsymbol{\xi})$:
\[
\nabla\cdot\left(\bar{\bar{\Xi}}\cdot\nabla\right)=\frac{1}{J}\partial_{i}(J\Xi^{ij}\partial_{j}),
\]
with $J$ the Jacobian of the transformation, $\partial_{i}=\partial/\partial\xi_{i}$,
and $\Xi^{ij}=\nabla\xi_{i}\cdot\bar{\bar{\Xi}}\cdot\nabla\xi_{j}$
the contravariant components of the tensor $\bar{\bar{\Xi}}$. Therefore,
the discrete treatment discussed below in Cartesian geometry can be
generalized readily by considering the curvilinear tensor components
$J\Xi^{ij}$ instead of the Cartesian ones, and dividing the discretization
formulas by the cell Jacobian.

Our starting point is a centered, conservative discretization of the
flux derivatives along a given coordinate direction, e.g.:

\begin{equation}
\frac{\partial F_{x}}{\partial x}|_{i,j}=\frac{\tilde{F}_{x,i+1/2,j}-\tilde{F}_{x,i-1/2,j}}{\mathit{\Delta x}},\label{eq:discrete-divergence}
\end{equation}
where $\tilde{F}_{x,i+1/2,j}$ and $\tilde{F}_{x,i-1/2,j}$ are fluxes
through $\left(i+\frac{1}{2},j\right)$ \ and $\left(i-\frac{1}{2},j\right)$
faces along the $x$-direction, which bound the cell $\left(i,j\right)$.
The definitions of the fluxes will determine the derivative's precision
as well as other aspects, such as positivity preservation. A fourth-order
centered discretization can be constructed from Eq. \ref{eq:discrete-divergence}
by interpolating the fluxes using four discrete points, two at each
side of the face, as:
\begin{equation}
\tilde{F}_{x,i+1/2,j}=\overset{2}{\underset{l=-1}{\sum}}a_{l}F_{x,i+l,j},\label{eq:4th-order-derivative}
\end{equation}
with $a_{-1}=a_{2}=-\frac{1}{12}$, $a_{0}=a_{1}=\frac{7}{12}$ .

\begin{figure}
\begin{centering}
\includegraphics{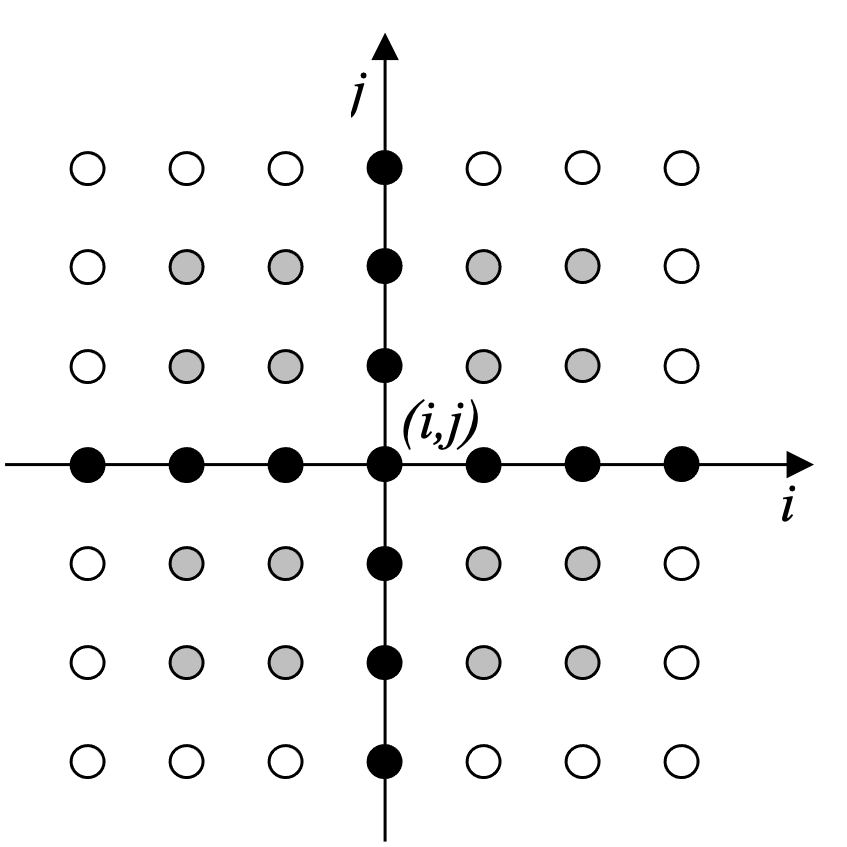}
\par\end{centering}
\caption{\protect\label{fig:global-stencil}Combined stencil for discretization
of parallel transport operator. Black points correspond to co-derivative
fluxes, and gray points to cross-derivative ones.}
\end{figure}

However, the flux itself in the operator of interest is given by $F_{x}=\Xi^{xx}\partial_{x}T+\Xi^{xy}\partial_{y}T$,
which must be estimated to fourth-order as well. According to Eq.
\ref{eq:4th-order-derivative}, we can write: 
\begin{eqnarray}
\frac{\partial}{\partial x}\left(\Xi^{xx}\frac{\partial f}{\partial x}\right)|_{i,j} & = & \frac{1}{\mathit{\Delta x}}\overset{2}{\underset{l=-1}{\sum}}a_{l}\left(\Xi_{i+l,j}^{xx}\frac{\partial f}{\partial x}|_{i+l,j}-\Xi_{i-1+l,j}^{xx}\frac{\partial f}{\partial x}|_{i-1+l,j}\right),\label{eq:4th-order-xx}\\
\frac{\partial}{\partial x}\left(\Xi^{xy}\frac{\partial f}{\partial y}\right)|_{i,j} & = & \frac{1}{\mathit{\Delta x}}\overset{2}{\underset{l=-1}{\sum}}a_{l}\left(\Xi_{i+l,j}^{xy}\frac{\partial f}{\partial y}|_{i+l,j}-\Xi_{i-1+l,j}^{xy}\frac{\partial f}{\partial y}|_{i-1+l,j}\right).\label{eq:4th-order-xy}
\end{eqnarray}
The derivatives $\frac{\partial f}{\partial x}$ and $\frac{\partial f}{\partial y}$
must be computed with at least fourth-order accuracy. Co-derivative
fluxes (that is, those with both derivatives along the same direction)
are calculated using a 7-point stencil. Cross-derivative fluxes are
computed using a $5\times5$-point stencil to fit a fourth-order polynomial.
Special treatment is required at physical boundaries to retain fourth-order
accuracy with a single ghost-cell layer. At logical boundaries between
parallel domains, we employ the same treatment as at physical boundaries
and then perform a synchronization step to average fluxes from left
and right domains to enforce strict conservation. This preserves conservation
and high-order accuracy without needing to communicate more than once
ghost-cell layer. The resulting 29-point 2D stencil for the calculations
of Eqs. \ref{eq:4th-order-xx} and \ref{eq:4th-order-xy} is illustrated
in Fig. \ref{fig:global-stencil}, and discussed in detail in \ref{app:discrete-details}.

\subsection{Reformulation and discretization of cross-derivative diffusion fluxes}

Cross-diffusion fluxes are problematic from a discretization and solver
robustness standpoint. At the discretization level, assuming a second-order
accurate discretization for the sake of argument, cross-derivative
fluxes are responsible for the lack of a maximum principle \citep{sharma2007preserving,dutoit2018positivity}.
Co-derivative ones are positivity-preserving for the standard second-order
star-stencil, but this property is in principle lost with a fourth-order implementation. Nevertheless,
we have found the scheme to be robust against developing negative temperature
values in our numerical tests.

Several authors \citep{sharma2007preserving,dutoit2018positivity,lipnikov2012minimal,gyrya2017arbitrary}
have explored the use of advective-like limiters for the cross-diffusion
terms to control positivity. Here, we follow Ref. \citep{dutoit2018positivity}
because of its generality, ease of implementation in 2D and 3D by
reusing existing advective routines, and amenability to multigrid
preconditioning. For exposition purposes, we consider the cross-diffusion
term $\text{\ensuremath{\partial_{x}}(\ensuremath{\Xi^{xy}}}\partial_{y}T)$.
In the reference, Du Toit and collaborators propose to reformulate
this term as:
\[
\text{\ensuremath{\partial_{x}}(\ensuremath{\Xi^{xy}}}\partial_{y}T)=\text{\ensuremath{\partial_{x}}}\left(T\Xi^{xy}\frac{\partial_{y}T}{T}\right)=\text{\ensuremath{\partial_{x}}}\left(Tv_{x}^{*}(T)\right),
\]
where $v_{x}^{*}(T)=\Xi^{xy}(\partial_{y}T)/T$ is a fictitious velocity.
The resulting advection operator is nonlinear in $T$ even if the
original cross-diffusion term is linear, but this is to be expected as we
are effectively introducing a nonlinear limiter to the cross-diffusion
term to enforce positivity. In practice, the fictitious velocity can
be found discretely at the face $i+1/2$ from the cross fluxes as
(from Eq. \ref{eq:4th-order-xy}):
\[
v_{x,i+1/2,j}^{*}=\frac{1}{T_{i+1/2,j}}\overset{2}{\underset{l=-1}{\sum}}a_{l}\left(\Xi_{i+l,j}^{xy}\frac{\partial f}{\partial y}|_{i+l,j}\right),
\]
where the flux inside the sum is computed as detailed in \ref{app:discrete-details},
and we estimate the face-centered temperature in the denominator by
a simple average of adjacent cell centers and adding a small failsafe
$\epsilon_{fs}=10^{-4}$ to prevent it from getting to close to zero,
i.e., $T_{i+1/2,j}=(T_{i,j}+T_{i+1,j})/2+\epsilon_{fs}$.

Once so transformed, any monotonicity-preserving advective discretization
scheme can in principle be used. However, most such schemes are designed
for explicit integrators, which are not suitable for our application
due to the numerical stiffness present from the large parallel-transport
coefficients. In an implicit context, one needs to be careful as many
advective limiters are highly nonlinear and not differentiable, which
is problematic for the Jacobian computation in Newton's method, and
therefore difficult to deal with in a nonlinear iterative solver context.
In this study, we use the QUICK/second-order-upwind variant of the
SMART scheme \citep{smart} for our advective discretization needs,
which is up to third-order accurate when possible, and reverts locally
to second- or first-order upwinding when needed to preserve monotonicity.
We have found the SMART variant to be robust, well behaved, and amenable
for JFNK nonlinear iterative solvers. We discuss our implicit
timestepping and nonlinear solver strategy next.

\subsection{Temporal discretization, nonlinear solver strategy, and multigrid
preconditioning}

For time discretization, we employ a second-order-accurate Backward
Differentiation scheme (BDF2) \citep{byrne-acmtms-75-bdf}. BDF2 is
not strictly positivity-preserving, but it provides a good balance
between accuracy and robustness in practice (as our numerical tests
will demonstrate).

As discussed earlier, the discretization of advectionalized cross-fluxes
is nonlinear. Additional sources of nonlinearity may originate from
the the transport coefficients, which may depend explicitly on the
dependent variable (see e.g. \citep{braginskii} for fusion plasmas).
This, combined with the need of implicit timestepping, imply a nonlinear
solver is needed. Here we consider JFNK \citep{knoll-jcp-04-nk_rev}
owing to its versatility and robustness in practice.

JFNK needs to be preconditioned for efficiency (details below). In
JFNK, the preconditioning step is naturally incorporated in the Krylov
method (flexible GMRES \citep{saad-fgmres}). Nonlinear convergence
is controlled in our implementation by the usual criterion:
\begin{equation}
\left\Vert \mathbf{F}(\mathbf{x}_{k})\right\Vert _{2}<\epsilon_{a}+\epsilon_{r}\left\Vert \mathbf{F}(\mathbf{x}_{0})\right\Vert _{2}=\epsilon_{t},\label{eq-Newton-conv-tol-1}
\end{equation}
where $\left\Vert \cdot\right\Vert _{2}$ is the $L_{2}$-norm (Euclidean
norm), $\epsilon_{a}=\sqrt{N}\times10^{-15}$ (with $N$ the total
number of degrees of freedom) is an absolute tolerance to avoid converging
below round-off, $\epsilon_{r}$ is the Newton relative convergence
tolerance, and $\mathbf{F}(\mathbf{x}_{0})$ is the initial residual.
We also have the option to control convergence by directly measuring the absolute size
of the Newton update:
\begin{equation}
\left\Vert \delta\mathbf{x}_{k}\right\Vert <\epsilon_{s}.\label{eq:sol-tol}
\end{equation}
By default, we set $\epsilon_s=0$ unless otherwise stated. We employ an inexact Newton method \citep{inexact-newton} in which
the FGMRES convergence tolerance is adjusted every Newton iteration
as follows:
\begin{equation}
\left\Vert J_{k}\delta\mathbf{x}_{k}+\mathbf{F}(\mathbf{x}_{k})\right\Vert _{2}<\zeta_{k}\left\Vert \mathbf{F}(\mathbf{x}_{k})\right\Vert _{2},\label{eq-inexact-newton-1}
\end{equation}
where $\zeta_{k}$ is the inexact Newton parameter and $J_{k}=\left.\frac{\partial\mathbf{F}}{\partial\mathbf{x}}\right|_{k}$
is the Jacobian matrix. Here, we employ the same prescription for
$\zeta_{k}$ as in earlier studies \citep{chacon-JCP-hall,chacon-pop-08-3dmhd}:
\begin{eqnarray*}
\zeta_{k}^{A} & = & \gamma\left(\frac{\left\Vert \mathbf{F}(\mathbf{x}_{k})\right\Vert _{2}}{\left\Vert \mathbf{F}(\mathbf{x}_{k-1})\right\Vert _{2}}\right)^{\alpha},\\
\zeta_{k}^{B} & = & \min[\zeta_{max},\max(\zeta_{k}^{A},\gamma\zeta_{k-1}^{\alpha})],\\
\zeta_{k} & = & \min[\zeta_{max},\max(\zeta_{k}^{B},\gamma\frac{\epsilon_{t}}{\left\Vert \mathbf{F}(\mathbf{x}_{k})\right\Vert _{2}})],
\end{eqnarray*}
with $\alpha=1.5$ , $\gamma=0.9$, and $\zeta_{max}=0.8$. The convergence
tolerance $\epsilon_{t}$ is the same as in Eq.~\ref{eq-Newton-conv-tol-1}.

For preconditioning, we employ a geometric MG solver. The MG preconditioner
inverts a low-order discretization of the heat transport equation
(i.e., second-order discretization of co-derivative fluxes and first-order
upwind discretization of advective ones, both linear), which is guaranteed
to be \emph{h}-elliptic and therefore to have the smoothing property
(and to lead to a successful MG method) \citep{brandt2011multigrid}.
The nonlinear dependence on temperature in the transport coefficients
(when present) and advective cross-fluxes is Picard-linearized in
the MG solve to the previous nonlinear iteration. Our MG implementation
features a matrix-light implementation \citep{chacon-JCP-hall,chacon-pop-08-3dmhd,chacon2024hall},
in which only the diagonal of the system of interest is stored for
smoothing purposes. Coarse operators are found via rediscretization,
and required residuals in the MG iteration are found in a matrix-free
manner. For smoothing, unless otherwise specified we employ five passes
of damped Jacobi (p. 10 in \citep{briggs-MG}; p. 118 in \citep{wessMG}),
with weight $\omega=0.7$, for both the restriction and the prolongation
steps. MG restriction employs conservative agglomeration, and prolongation
employs a first-order interpolation. The coarsest-mesh problem (on
an $8\times8$ mesh) is solved with GMRES. In MG jargon, such V-cycle
is identified as V(5,5), where the two integers indicate restriction
and prolongation smoothing steps, respectively.

\section{Numerical tests}

\label{sec:numtests}We present several numerical tests that will
demonstrate the convergence and performance properties of the fourth-order
Eulerian anisotropic transport implicit algorithm. The tests will
also highlight the physical need for retaining high anisotropy in
the heat flux. To characterize the test results, we consider units where the length scale is
the domain size $L$ and the time scale is based on the perpendicular
thermal diffusivity, $\tau_{r}=L^{2}/\chi_{\perp}$. For constant
transport coefficients, the heat transport equation (Eq. \ref{diffusion_eq})
becomes, 
\begin{equation}
\frac{\partial T}{\partial\hat{t}}-\frac{\chi_{\parallel}}{\chi_{\perp}}\hat{\nabla}_{\parallel}^{2}T-\hat{\nabla}_{\perp}^{2}T=S.\label{eq:dimless-T-eq}
\end{equation}
After temporal discretization, two key parameters characterize the
numerical challenges in Eq. \ref{eq:dimless-T-eq}. The first one
is $\Delta\hat{t}\chi_{\parallel}/\chi_{\perp}=\Delta t\chi_{\parallel}/L^{2}$,
which is independent of $\chi_{\perp}$. The second one is a dimensionless
measure of the anisotropy, which can be defined as:
\begin{equation}
\epsilon=\frac{\hat{\tau}_{\parallel}}{\hat{\tau}_{\perp}},\label{eps}
\end{equation}
where $\hat{\tau}_{\parallel}$, $\hat{\tau}_{\perp}$are fast and
slow timescales for the parallel and perpendicular diffusivities,
respectively, and are given as \citep{chacon2014asymptotic}: 
\begin{equation}
\hat{\tau}_{\parallel}=\frac{\tau_{\parallel}}{\tau_{r}}=\frac{L_{\parallel}^{2}}{\chi_{\parallel}}\frac{\chi_{\perp}}{L^{2}}=\frac{\chi_{\perp}}{\chi_{\parallel}}\hat{L}_{\parallel}^{2}\,,\label{timescale1}
\end{equation}

\begin{equation}
\hat{\tau}_{\perp}=\frac{\tau_{\perp}}{\tau_{r}}=\frac{L_{\perp}^{2}}{\chi_{\perp}}\frac{\chi_{\perp}}{L^{2}}=\hat{L}_{\perp}^{2}\,.\label{timescale2}
\end{equation}
Here, $L_\parallel$, $L_\perp$ are parallel and perpendicular temperature-gradient length scales, respectively. Both $\chi_\parallel \Delta t$ and $\epsilon$ will be useful to characterize
the numerical results below.

\subsection{NIMROD benchmark test}

\label{sec:nimrod} The NIMROD benchmark
test \citep{sovinec2004nonlinear} will allow us to evaluate
the numerical pollution for both second- and fourth-order schemes,
and verify that they possess the expected scalings as the grid resolution
is increased. The comparison between the methods will demonstrate
the significant improvement in numerical pollution with the fourth-order
method. In addition, the NIMROD benchmark test will allow us to evaluate
the performance of the solver as the anisotropy of the heat flux is
increased.

The two-dimensional NIMROD test has a heat source $S(x,y)=-\nabla^{2}\psi(x,y)$
and a magnetic field $\mathbf{B}=\mathbf{z}\times\nabla\psi$, where
the flux function is initialized as $\psi(x,y)=cos(\pi x)cos(\pi y)$
in the domain $(x,y)\,\epsilon\,(-0.5,0.5)\times(-0.5,0.5)$. The
temperature has homogeneous Dirichlet boundary conditions at all boundaries.
The flux function is time-independent, and therefore the magnetic
field is fixed in time. Since $\psi$ is an eigenmode of the Laplacian
for the domain and boundary conditions chosen, the temperature remains
a flux function at all times and therefore should remain in the null
space for the $\nabla_{\parallel}$ operator, $\mathcal{N}(\nabla_{\parallel})$.
The temperature analytical solution is: 
\begin{equation}
T(x,y,t)=\frac{1-\exp\left(-2\chi_{\perp}\pi^{2}t\right)}{\chi_{\perp}}\psi(x,y)\,,\label{nimrodtemp}
\end{equation}
with $T(0,0,t\rightarrow\infty)=1/\chi_{\perp}$.

Following \citep{sovinec2004nonlinear}, a convenient measure of perpendicular
numerical pollution is: 
\begin{equation}
\Delta\chi=\frac{1}{T(0,0,t\rightarrow\infty)}-\chi_{\perp}\,,\label{nimrodsteadystate}
\end{equation}
where $\chi_{\perp}$ is an input parameter and $T(0,0,t\rightarrow\infty)$
is computed numerically. By design, both parallel and perpendicular
length scales are of $\mathcal{O}(1)$, so the anisotropy parameter
is simply the ratio of diffusivity coefficients (which are constant
in this test), $\epsilon=\chi_{\perp}/\chi_{\parallel}$. This test
considers various $\epsilon$-values for both second- and fourth-order
schemes at uniform grid resolutions of 32$\times$32, 64$\times$64
and 128$\times$128.

\begin{figure}
\centering{}\includegraphics[width=0.5\columnwidth]{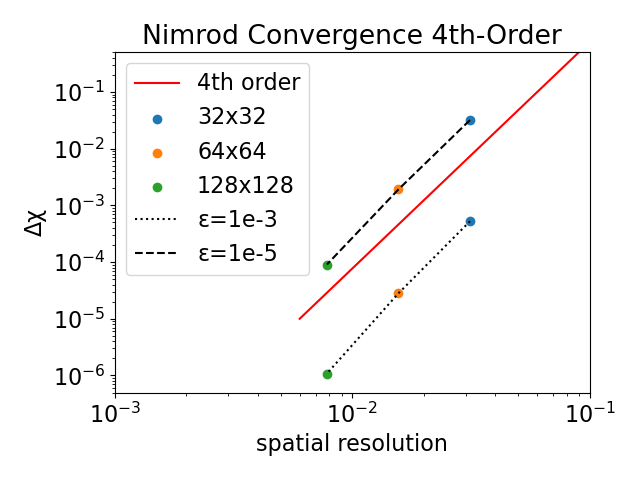}\includegraphics[width=0.5\columnwidth]{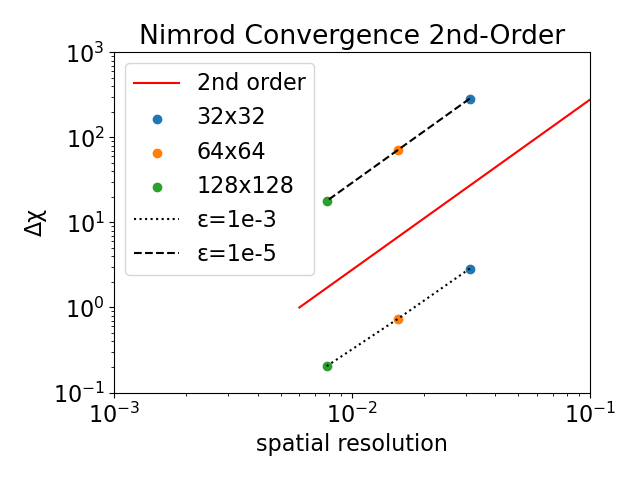}\caption{NIMROD benchmark test: relative error $\Delta\chi$ at steady state
for second-order and fourth-order discretizations with various grid
resolutions and $\epsilon=10^{-3},10^{-5}$. Expected asymptotic convergence
rates are recovered by the scheme.\protect\label{nimrodconv}}
\end{figure}

\begin{figure}
\centering{}\includegraphics[scale=0.5]{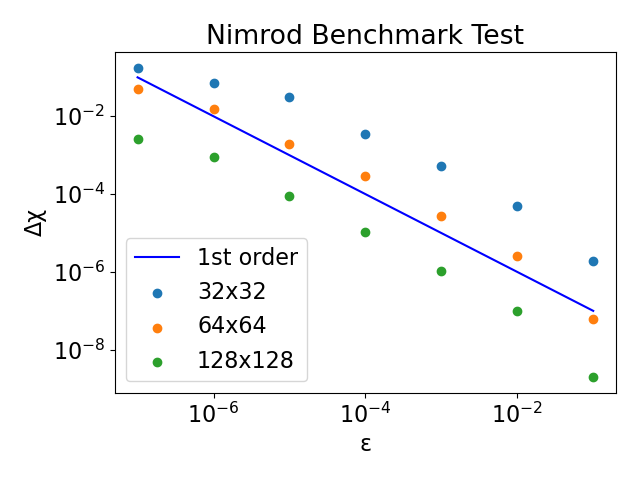} \caption{NIMROD benchmark test: scaling of $\Delta\chi$ with the anisotropy
$\epsilon$. \protect\label{nimroderr}}
\end{figure}

We consider accuracy first. Fig. \ref{nimrodconv} shows that the
grid convergence rates for both second- and fourth-order discretizations scale as expected.
The numerical pollution can be seen to be greatly reduced by the fourth-order
scheme compared to the second-order one, by up to five orders of magnitude.
For the second-order method, $\Delta\chi\gg1$, which makes it unusable
in practice as a solver (but performs well as a preconditioner, as
we shall see). 
 Finally, Fig. \ref{nimroderr} shows that the numerical pollution
error scales as $1/\epsilon$ (i.e., the method is not asymptotic-preserving).
This is expected behavior whenever the discretization does not have
a mesh point exactly at the O-point \citep{van2016finite} (which
is the case here), since the source of the numerical error is the
$\frac{1}{\epsilon}\nabla_{\parallel}^{2}T$ term. The magnitude of
the error is manageable for the $\epsilon$-values considered owing
to the fourth-order accuracy of the scheme.

\begin{figure}
\centering{}\includegraphics[width=0.5\columnwidth]{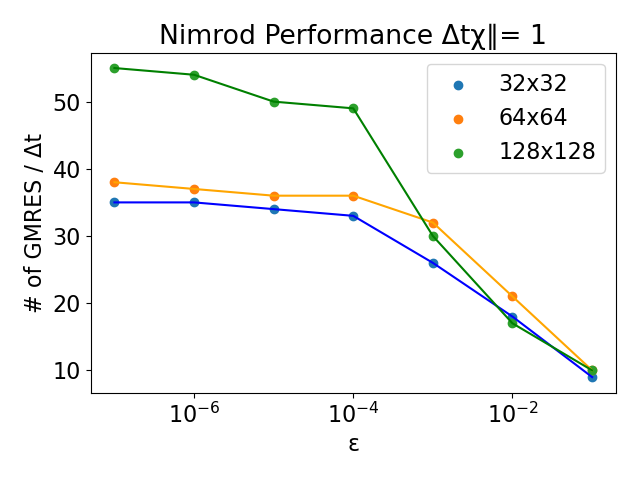}\includegraphics[width=0.5\columnwidth]{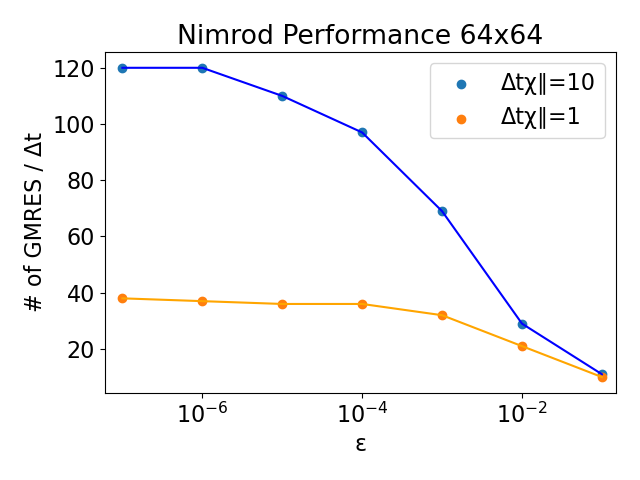}\caption{NIMROD benchmark test: number of GMRES iterations per time step vs.
$\epsilon$ for the fourth-order scheme for (a) different mesh resolutions
for $\Delta t\chi_{\parallel}=1$, and (b) various $\Delta t\chi_{\parallel}$
for the $64\times64$ mesh. \protect\label{fig:nimrodperf}}
\end{figure}

In regards to performance, Fig. \ref{fig:nimrodperf}-a depicts the
results of a grid-convergence study as a function of $\epsilon$ for
$\Delta t\chi_{\parallel}=1$. Performance is measured as the average
number of GMRES iterations per time step for a relative JFNK nonlinear
tolerance of $\epsilon_{r}=10^{-3}$ . The results demonstrate that
the fourth-order method can handle very large anisotropies with minor
performance degradation under mesh refinement (an increase of 16$\times$
in the total number of mesh points only results in a 1.5$\times$
increase in GMRES iterations). Also, performance of the solver is
independent of $\epsilon$ for sufficiently small values. Fig. \ref{fig:nimrodperf}-b
shows performance sensitivity with timestep for various $\epsilon$,
suggesting that the average number of GMRES iterations per timestep
scales as $\sim\sqrt{\Delta t\chi_{\parallel}}$. This scaling favors
the use of larger timesteps for efficiency (as allowed by the dynamics
of interest), since the overall simulation cost scales as $\sim N_{\Delta t}\sqrt{\Delta t}=(t_{final}/\Delta t)\times\sqrt{\Delta t}\sim t_{final}/\sqrt{\Delta t}$.

\subsection{Magnetic island test}

\label{sec:island} This test is inspired by a fusion-relevant magnetically confined plasma, and was originally proposed in Ref. \citep{gunter2009mixed}.
It features a more interesting B-field topology in a 2D cylindrical
geometry $(r,\theta)$. The magnetic field is defined by the flux
function:
\begin{equation}
\psi(r,\theta)=(r-r_{s})^{2}+\delta r^{2}(1-r^{4})cos(\theta)\,,\label{islandpsi}
\end{equation}
in the domain $(r,\theta)\,\in\,[0,1]\times[0,2\pi)$. The island
width is $\delta=0.005$, with the island located at the rational
surface $r_{s}=0.7$. The magnetic field is given by $\mathbf{B}=\mathbf{z}\times\nabla\psi+\mathbf{B}_{z}$
with a guide field $B_{z}=1$, which is much stronger than the poloidal
field at the location of the island. As in the NIMROD test, the magnetic
field is constant in time.

\begin{figure}
\centering{}\includegraphics[width=0.5\columnwidth]{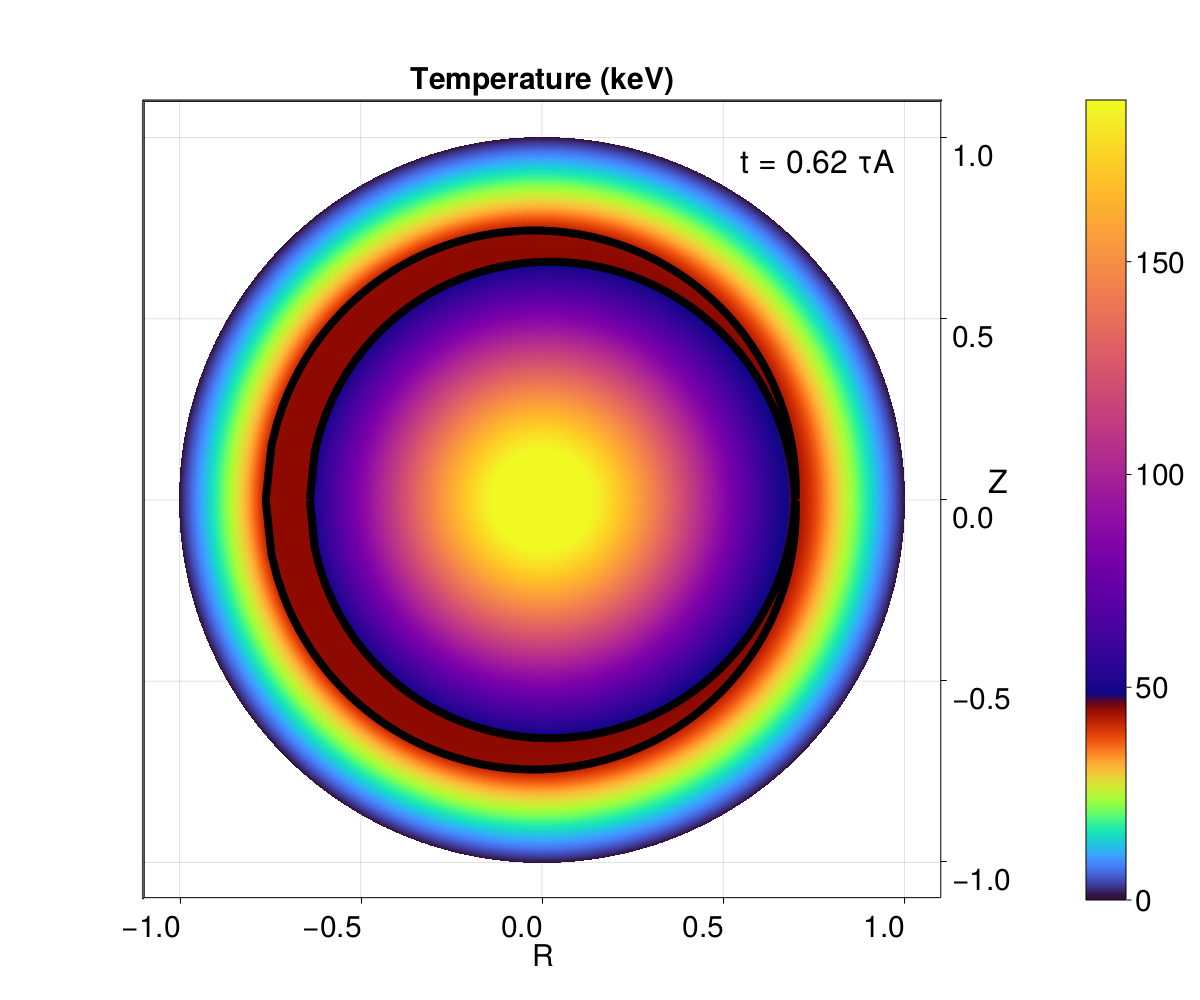}\caption{Magnetic island test: steady-state solution (t = 0.62) in Cartesian
coordinates for $\chi_{\perp}/\chi_{\parallel}=10^{-7}$ on a 512x256
mesh. The magnetic island separatrix is indicated with a black line.\protect\label{fig:islandsteady}}
\end{figure}

A time-independent heat source $S(x,y)=4(1-r^{2})^{8}$ is prescribed.
The initial condition is $T(\mathbf{x},t=0)=0$. Temperature boundary
conditions are homogeneous Dirichlet at $r=1$ and regularity at $r=0$.
Since the initial temperature is zero, this test is useful to assess
robustness against the development of negative temperatures. The steady-state
temperature is reached at $t=0.62$, and is depicted in Fig.~\ref{fig:islandsteady}.
This simulation employed $\Delta t=10^{-4}$, $\chi_{\parallel}=10^{7}$
and $\chi_{\perp}=1$ (both uniform) on a 512x256 mesh. 
 The flattening of the temperature field within the island is apparent.

\begin{figure}
\centering{}\subfloat[\label{islandtemp1}$t=0.03$]{\includegraphics[width=0.5\columnwidth]{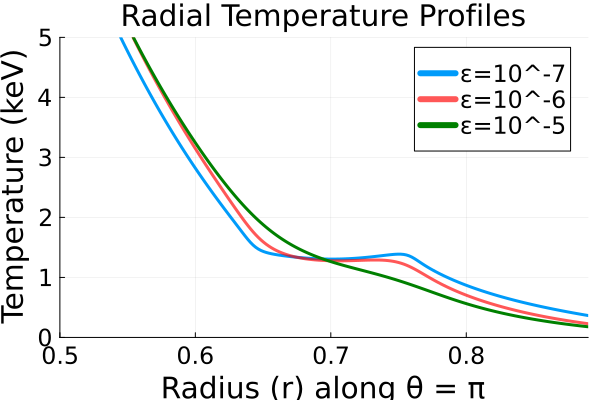}

}~\subfloat[\label{islandtemp2}$t=0.62$]{\includegraphics[width=0.5\columnwidth]{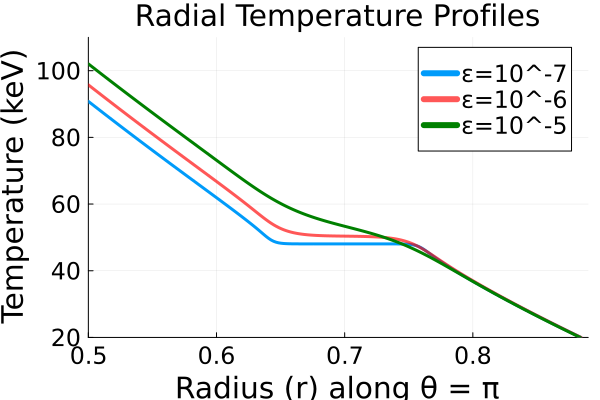}

}\caption{Magnetic island test: temperature profiles across the island at early
and late times. At large anisotropy, the flatness of the temperature
profile is preserved, whereas lower anisotropies have enough cross-field
diffusion to obscure the island's impact on the temperature profile.\protect\label{islandtemp}}
\end{figure}

Fig. \ref{islandtemp} shows radial temperature profiles across the
island for various anisotropies, both at early and late times. 
 At early times, the large anisotropy case ($\chi_{\parallel}=10^{7}$)
retains the flatness of the temperature profile within the island,
but this feature is progressively lost as the anisotropy is reduced.
The same is true at late times, but the accumulated heat-flux restriction
in the high-anisotropy case leads to a substantial decrease in the
final temperature at the core. This result highlights the benefits
of using as realistic an anisotropy ratio as the discretization allows
to capture physical temperature profiles.

\begin{figure}
\centering{}\includegraphics[width=0.5\columnwidth]{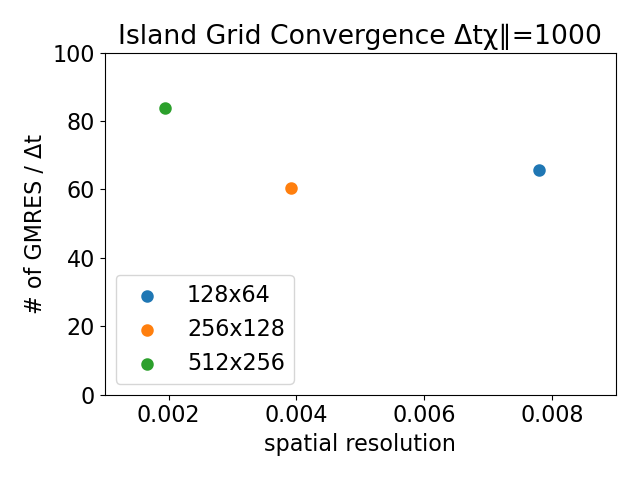}\includegraphics[width=0.5\columnwidth]{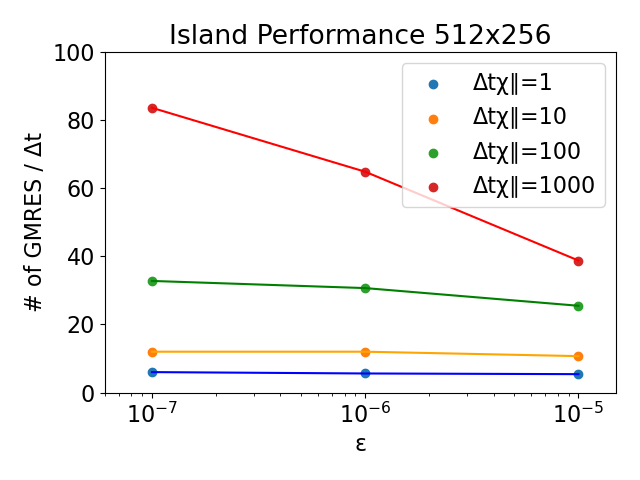}\caption{Magnetic island test. Left: Average number of GMRES iterations per
time step for various grid resolutions for the island test with $\chi_{\parallel}=10^{7}$
and a time step $\Delta t=10^{3}/\chi_{\parallel}=10^{-4}$ showing
no clear iteration growth with mesh refinement. Right: Average number
of GMRES iterations vs anisotropy $\epsilon\sim\chi_{\perp}/\chi_{\parallel}$
for a 512x256 mesh and various $\Delta t\chi_{\parallel}$ values.
\protect\label{fig:islandconv}}
\end{figure}

The magnetic island test is particularly challenging from a solver-performance
standpoint. We measure solver performance for a (very large) timestep $\chi_{\parallel} \Delta t=10^{3}$
in terms of the average number of iterations per time step (over the
whole simulation until steady state) to achieve a relative tolerance
of $5\times10^{-3}$ or a solution update tolerance of $\epsilon_{s}=10^{-9}$
(Eq. \ref{eq:sol-tol}), whichever comes first. First, we consider
a grid-convergence study with $\chi_{\parallel}/\chi_{\perp}=10^{7}$,
with results plotted in Fig. \ref{fig:islandconv}-left. 
 The results show no clear dependence of solver performance on the
mesh resolution, demonstrating that our second-order MG preconditioner
is effective in controlling the iteration count and in rendering the
solver almost optimal. The dependence of solver performance on $\epsilon$
(with fixed mesh resolution of $512\times256$) is depicted in Fig.
\ref{fig:islandconv}-right. The iteration count plateaus at high
anisotropy, just as for the NIMROD test in Fig. \ref{fig:nimrodperf},
and the results are consistent with a $\sqrt{\Delta t\chi_{\parallel}}$
scaling for small enough $\epsilon$.

\subsection{Bennett screw-pinch kink instability MHD simulation in 2D helical geometry}

\label{sec:helical}
\begin{figure}
\centering{}\includegraphics[width=3in]{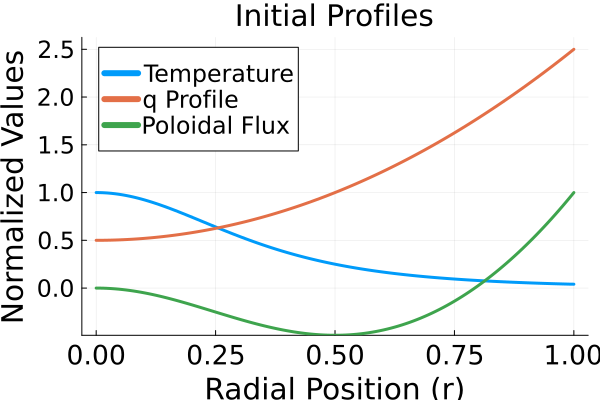} \caption{Bennett pinch test: Radial profiles of equilibrium temperature, magnetic
flux, and safety factor. The temperature and magnetic
flux are normalized to their maximum values.\protect\label{initialprofiles} }
\end{figure}

The third and fourth tests showcase the performance of the
scheme in a dynamic magnetohydrodynamics (MHD) simulation using the
PIXIE3D MHD code \citep{chacon-cpc-04-mhd_discret,chacon-pop-08-3dmhd}.
In this test, we initialize the simulation with a Bennett pinch equilibrium in cylindrical
geometry \citep{bennett1934magnetically}, which only depends on the
radial coordinate. The temperature is initialized as $T(r,t=0)=T_{0}/4(1+(4r^{2}))^{2}$,
with plasma beta $\beta=1$ and a uniform density. The field components
are initialized as $B_{\theta}(r,t=0)=2rB_{0}/(1+4r^{2})$ and $B_{y}(r,t=0)=B_{0}$.
The temperature $T$, poloidal magnetic flux $\Psi$, and safety factor
$q=rB_{z}/RB_{\theta}$ are depicted in Fig. \ref{initialprofiles}.
 The instability is seeded at the $q=1$ rational surface, located
at $r=0.5$. This setup is unstable to a kink instability, resulting
in a full magnetic topology reorganization via magnetic reconnection.
This reconnection event will allow fast parallel heat transport across
the newly formed flux surfaces only after the reconnection event.
Such a dramatic display is only possible when the code can handle
large heat-transport anisotropies correctly, or else perpendicular
numerical pollution will allow the energy to leak across flux surfaces
instead of flowing along them.

For the simulation, we consider a non-orthogonal 2D helical projection
of the 3D cylinder with domain $(r,\theta)\,\in\,[0,1]\times[0,2\pi)$
and helical pitch $m/n=1$, with $m$, $n$ the mode numbers in the
poloidal and axial directions. Boundary conditions are as follows:
regularity at $r=0$ for all quantities; at $r=1$ we have ideal wall
for the magnetic field, homogeneous Neumann for the temperature, and
zero-flow for all velocity components; and periodic boundary conditions
in $\theta$. Further details of the MHD equations, the 2D helical
geometry definition, and the boundary conditions can be found in Ref.
\citep{chacon-cpc-04-mhd_discret} (with an update on the treatment
of regularity conditions at $r=0$ in the Appendix of Ref. \citep{delzanno2008electrostatic}).
We consider realistic functional dependencies for resistivity (Spitzer)
and heat-transport coefficients (Braginskii) \citep{braginskii} with
temperature $T$ and density $\rho$, given in Alfv\'enic dimensionless
units as:
\[
\eta=\eta_{0}\left(\frac{T_{e0}}{T_{e}}\right)^{\frac{3}{2}}\quad,\quad\chi_{\perp}=\chi_{\perp0}\sqrt{\frac{T_{e0}}{T_{e}}}\frac{\rho^{2}}{\rho_{0}^{2}}\frac{B_{0}^{2}}{B^{2}}\quad,\quad\chi_{\parallel}=\chi_{\parallel0}\left(\frac{T_{e}}{T_{e0}}\right)^{\frac{5}{2}}.
\]
Here, $T_{e0}$, $\rho_{0}$ and $B_{0}$ are the reference temperature,
density and magnetic-field magnitude, respectively. The dimensionless
reference constants are given by $\eta_{0}=5\times10^{-4}$, $\chi_{\perp,0}=5\times10^{-4}$
and $\chi_{\parallel,0}=10^{3}$. The maximum heat-transport anisotropy
ratio is $\chi_{\parallel,0}/\chi_{\perp,0}=2\times10^{6}$. We also
employ a constant ion viscosity $\nu_0=5 \times 10^{-4}$ in the momentum equation, and a constant particle diffusivity $D_0=5\times10^{-4}$ in the continuity equation
 to ensure smoothness of the density field around
the coordinate singularity. We consider a $128\times64$ mesh. We
employ an implicit timestep of $\Delta t=10^{-2}\tau_{A}$ (with $\tau_{A}=L/v_{A}$
the Alfv\'en time, and $v_{A}=B_{0}/\sqrt{\rho_{0}\mu_{0}}$ the
Alf\'en speed). This timestep corresponds to $\Delta t\chi_{\parallel}=10$.
The equilibrium is perturbed at $t=0$ with a small radial velocity,
$V_{r}(r,\theta,t=0)=10^{-3}\sin(\pi r)\cos{\theta}$.

\begin{figure}
\begin{centering}
\includegraphics[width=3.1in]{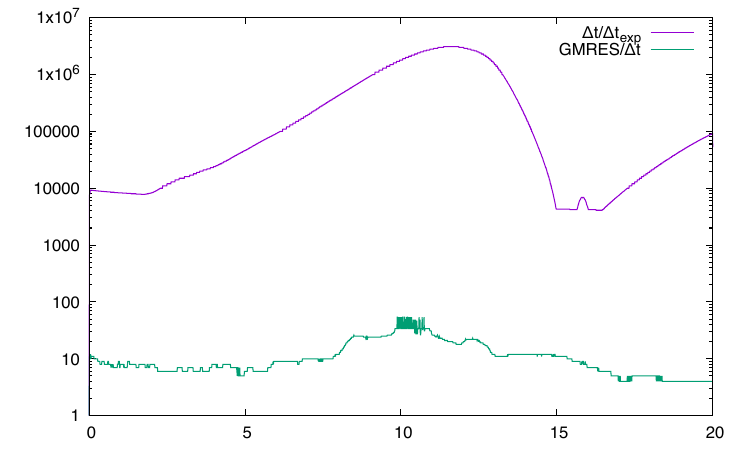}~\includegraphics[width=3.1in]{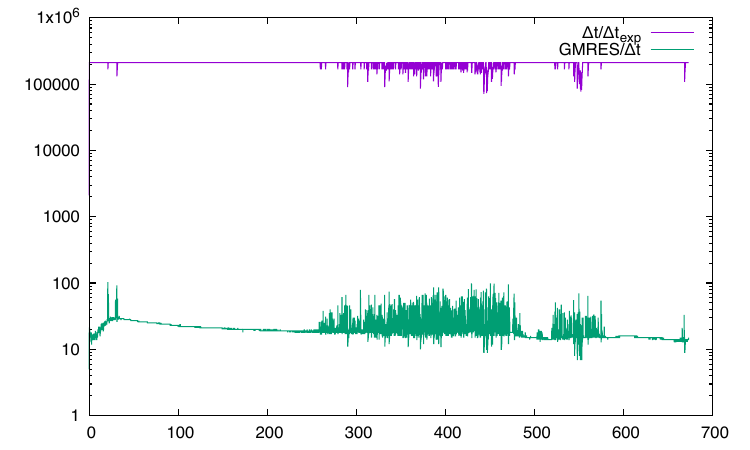}
\par\end{centering}
\caption{\protect\label{fig:cfl-history}Time history
of the $\Delta t/\Delta t_{exp}$ ratio and the number of GMRES iterations
per timestep for the Bennett test (left), and the ITER test (right).}
\end{figure}

At the beginning of the simulation, the implicit timestep is about
9300$\times$ larger than the explicit stability limit, $\Delta t_{exp}$, and peaks at $\sim 3.1\times10^{6} \Delta t_{exp}$
during the simulation. We run the simulation until $t=20\tau_{A}$.
For this test, the relative JFNK tolerance is $\epsilon_{r}=10^{-3}$,
and we use V(5,5) MG V-cycles for all the MG solves in the MHD preconditioner,
including the transport equation \citep{chacon-pop-08-3dmhd}. The
average number of GMRES iterations per timestep over the whole simulation
for the $128\times64$ mesh resolution is 12.0. The iteration and
$\Delta t/\Delta t_{exp}$ histories are provided in Fig. \ref{fig:cfl-history}-left,
demonstrating excellent algorithmic performance of the MHD solver
with the strongly anisotropic transport physics throughout the simulation
despite the very large implicit timestep used.

\begin{figure}
\centering{}\includegraphics[scale=0.17]{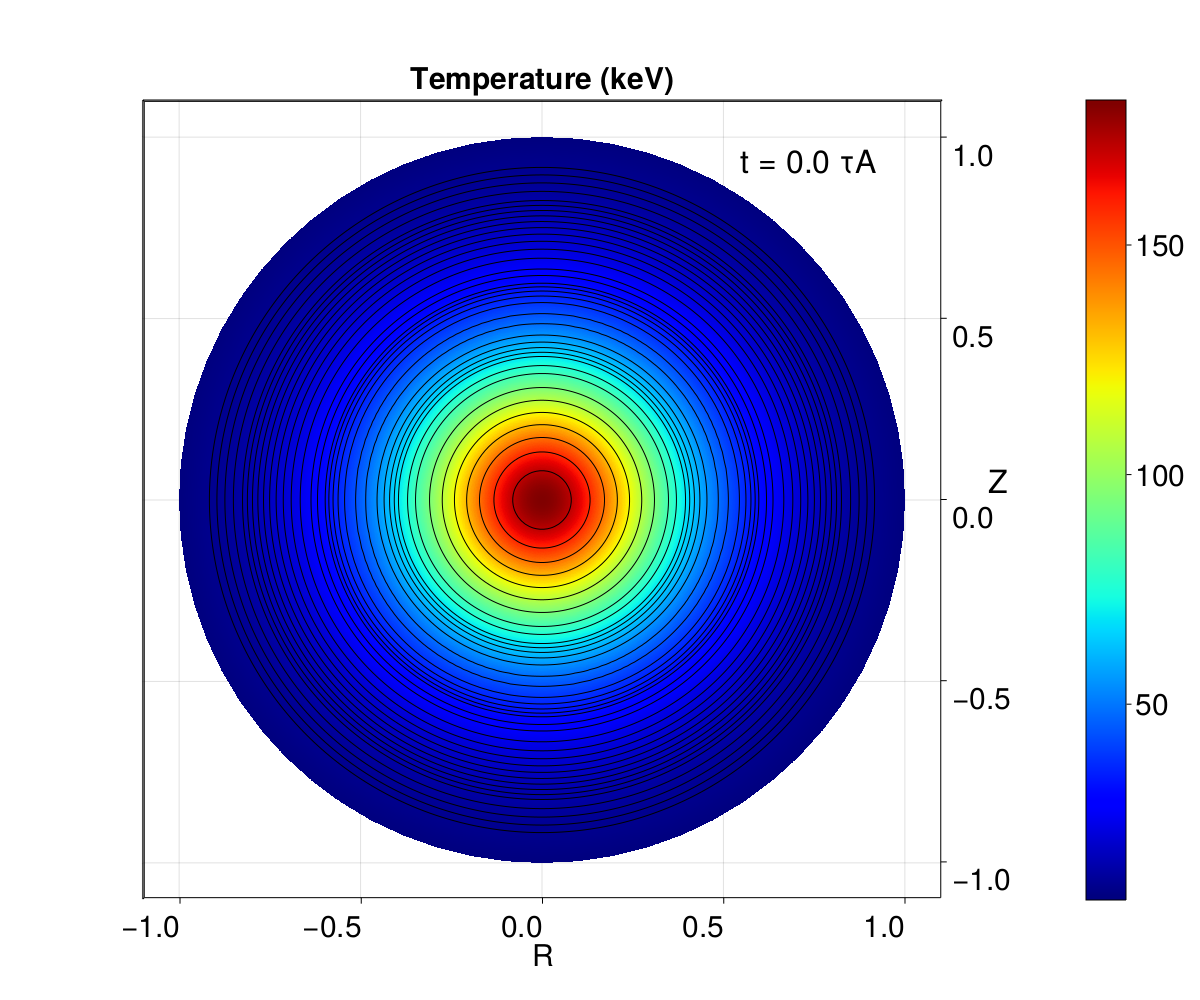}~\includegraphics[scale=0.17]{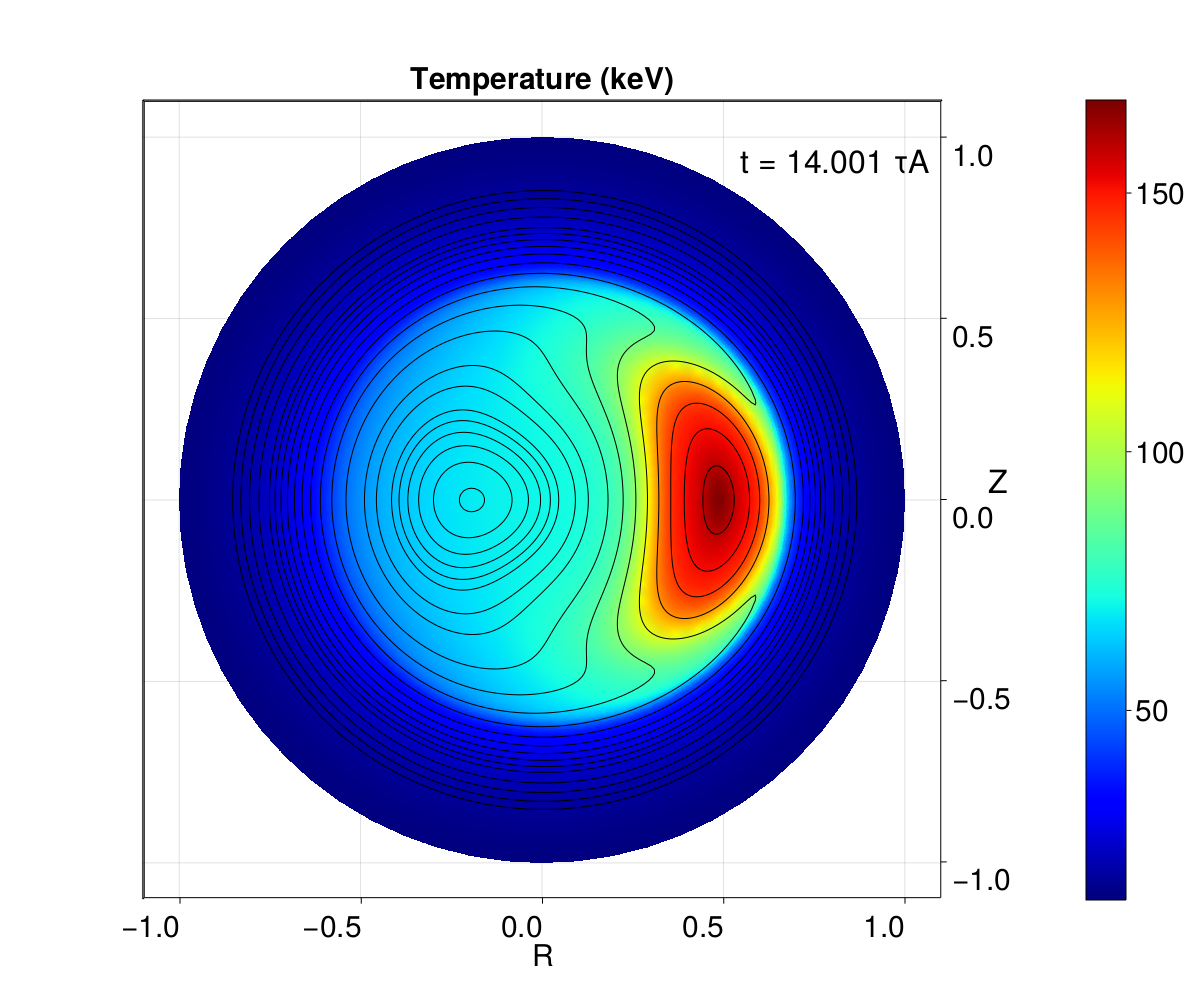}\\
 \includegraphics[scale=0.17]{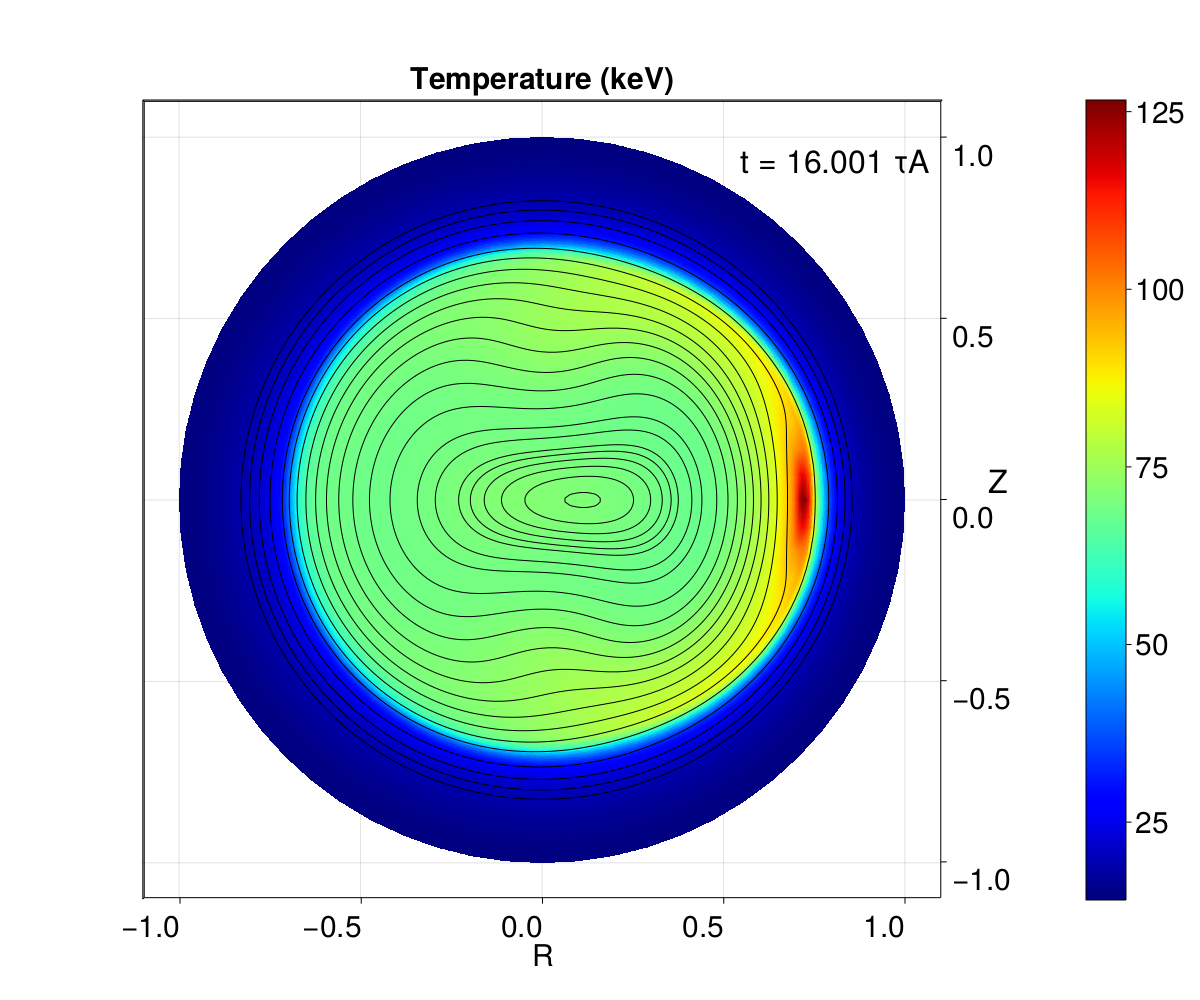}~\includegraphics[scale=0.17]{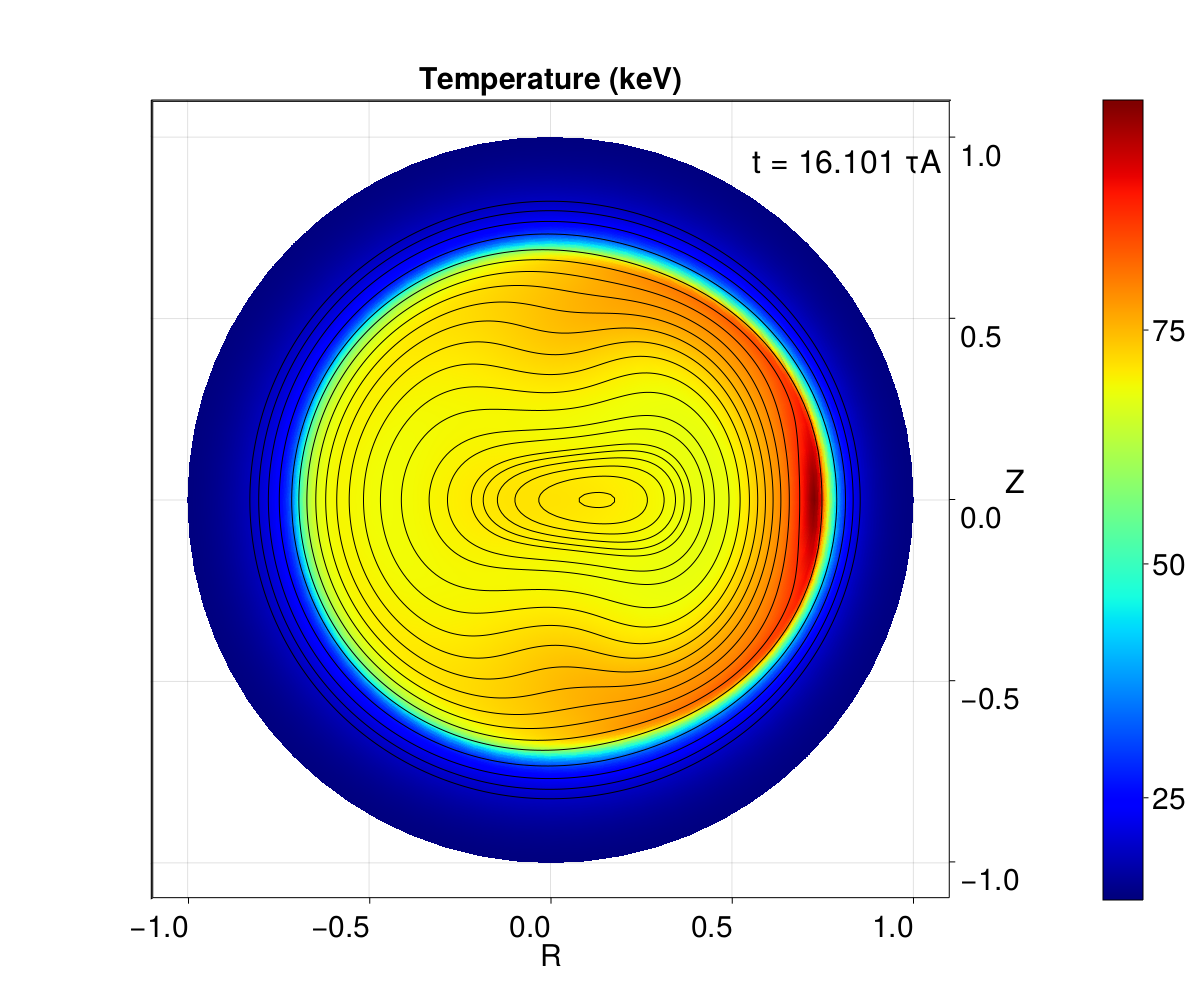}\\
 \includegraphics[scale=0.17]{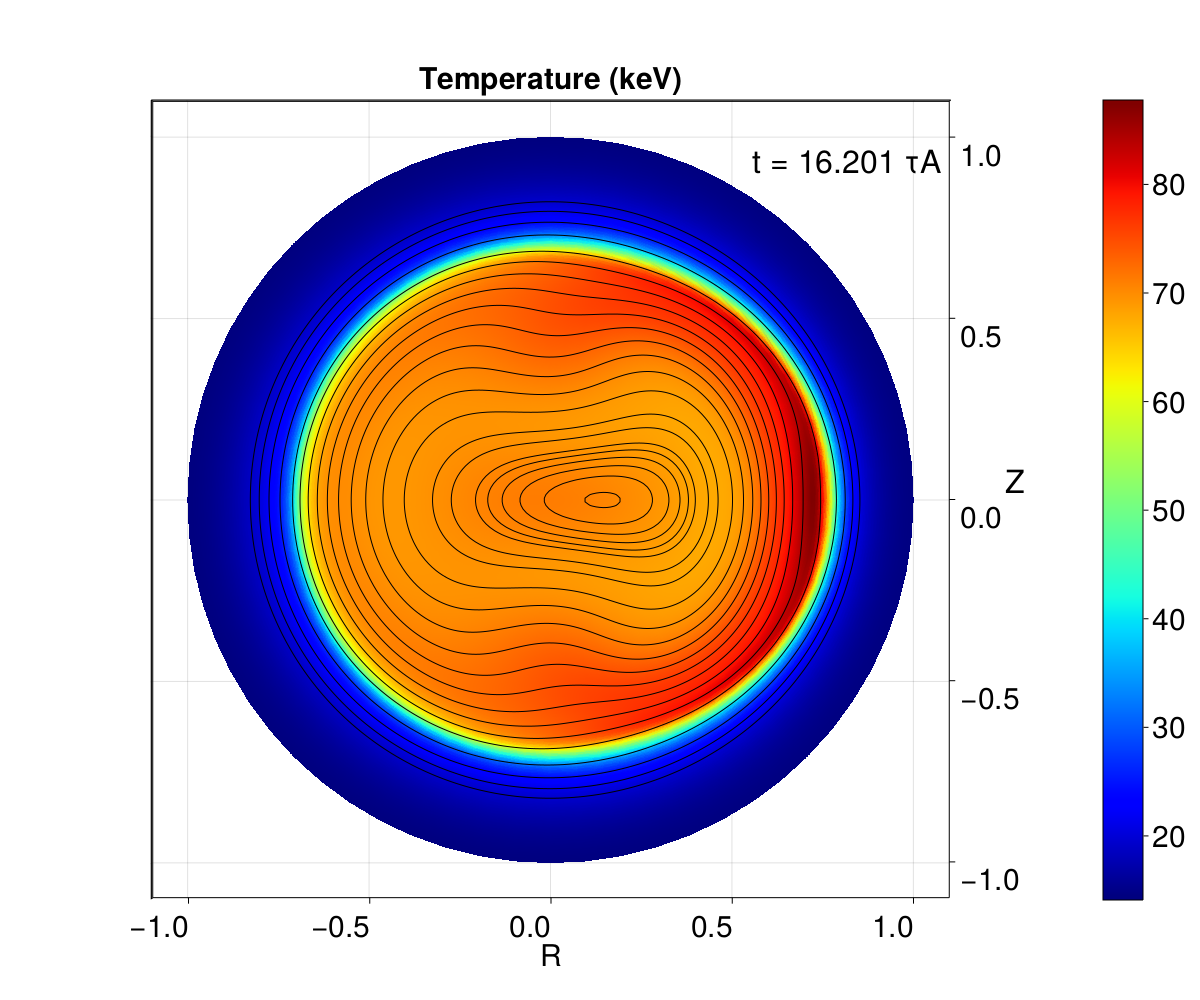}~\includegraphics[scale=0.17]{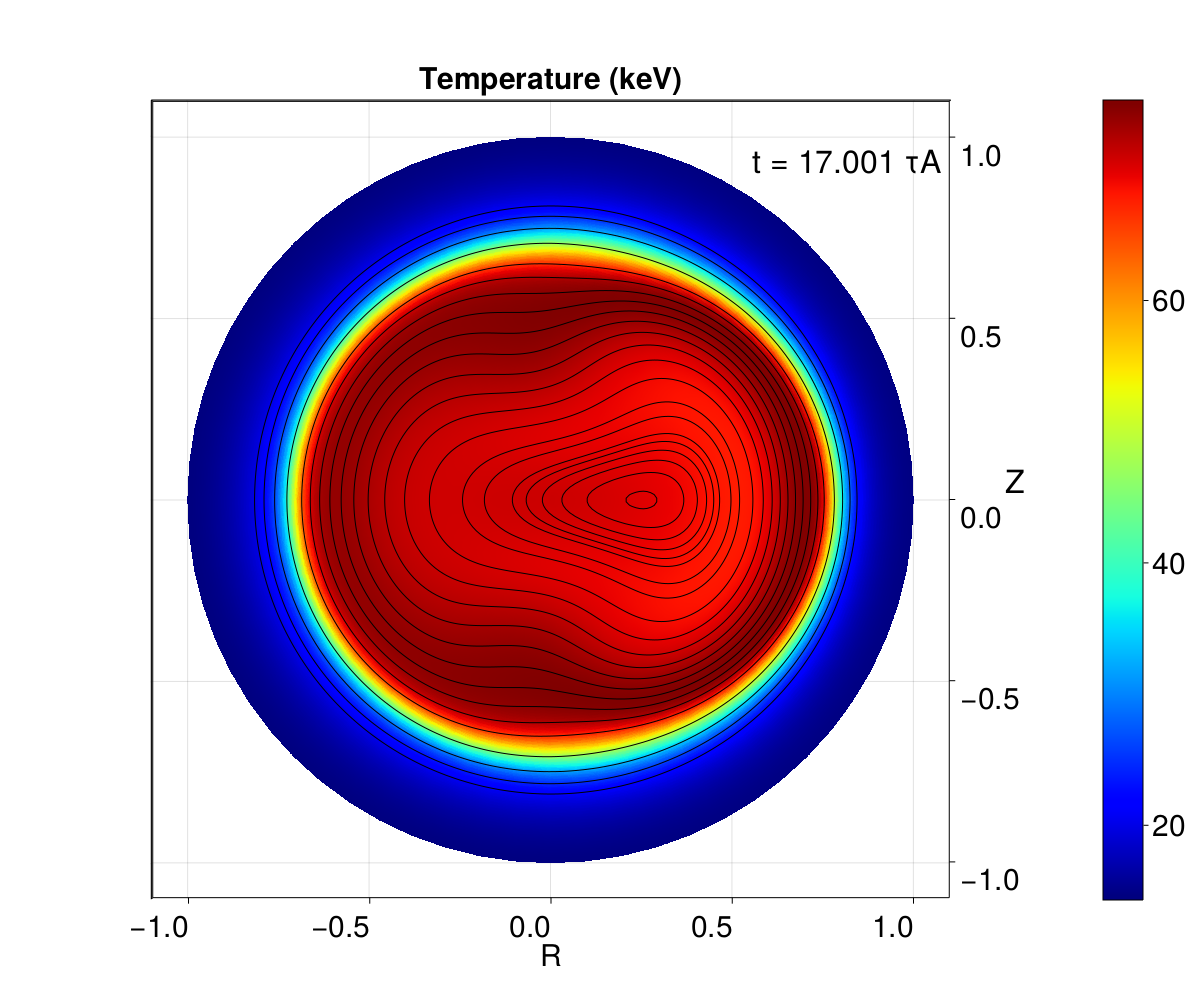}\caption{Bennett pinch test: Temperature snapshots of the closed flux surfaces
for the helical test at a resolution of $128\times64$. Timestamps provided in the figures. Note that
the color bar rescales with the maximum temperature at each snapshot.\protect\label{fig:helical32}}
\end{figure}

Snapshots of the temperature profiles overlayed with the flux surfaces
are shown in Fig. \ref{fig:helical32}. 
The fast reconnection event occurs around t = 16.1. Before the event,
most of the thermal energy is contained within the flux surfaces near
the X-point. Quickly after the reconnection event, this energy can
be seen to rapidly spread along the flux surfaces and fill the volume.
The sharpness of the results in this MHD simulation demonstrates the
ability of the scheme to perform accurately in highly coupled, nonlinear
MHD simulations on non-trivial geometries including coordinate singularities
(at $r=0$) and non-orthogonal geometries (helical).

\subsection{ITER kink instability MHD simulation in 3D toroidal geometry} 
\label{sec:iter}

\begin{figure}
    \centering
    \includegraphics[height=2.2in]{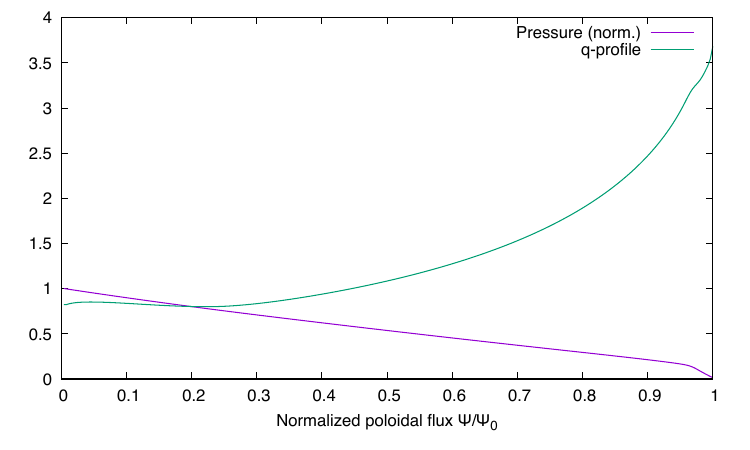}
    \includegraphics[height=2.15in,width=1.4in]{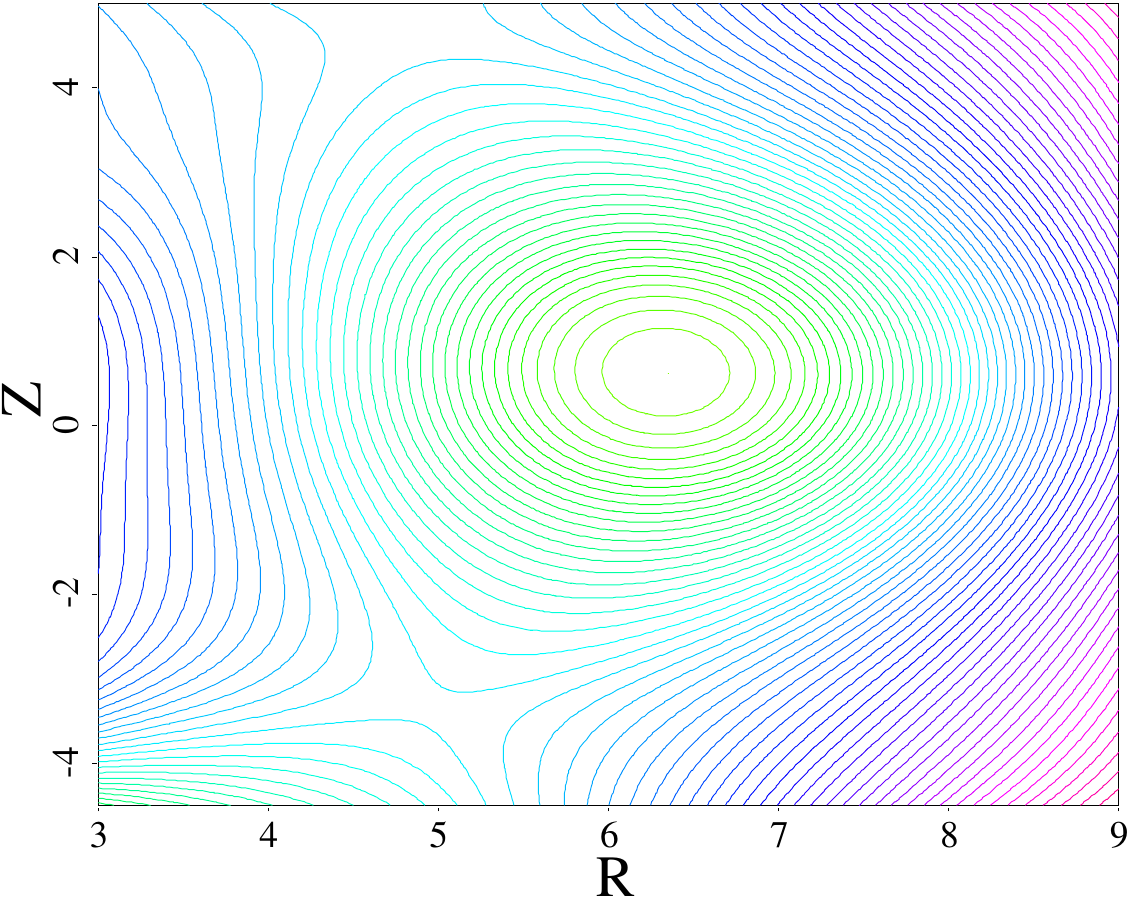}
    \caption{ITER test. Left: Profiles of initial pressure,  and q-profile as a function of the poloidal flux, normalized to the separatrix flux value. The pressure is normalized to the plasma beta, $\beta=0.028$. Right: ITER equilibrium poloidal flux.}\label{fig:initialprofiles2} 
\end{figure}

For our last test, we consider a fully 3D ITER tokamak MHD simulation of an unstable equilibrium in realistic non-orthogonal toroidal geometry. The equilibrium is unstable to a $m=1$, $n=1$ internal kink mode.
The initial pressure and safety-factor profiles are shown in Fig. \ref{fig:initialprofiles2}-left; the poloidal flux is provided in Fig. \ref{fig:initialprofiles2}-right.
Spitzer resistivity and Braginskii heat-transport coefficients are used as in the previous test, with  $\eta_0 = 10^{-4}$, $\chi_{\perp,0} = 2\times10^{-6}$, $\chi_{\parallel,0} = 20.0$ (seven orders of magnitude in thermal transport anisotropy). Additionally, we use $D_0 = 5 \times10^{-4}$ and $\nu_0 = 10^{-4} \abs{1+4.6 r^{3}}^4$, where $r$ is the radial logical coordinate.
The logical-to-physical coordinate map used for this simulation is described in \ref{app:iter-mesh}.
The increased viscosity at the edge of the plasma prevents generation of spurious oscillations in the cold plasma near the wall.
With seven orders of magnitude of transport anisotropy, thermal energy rapidly equalizes along field lines with very little diffusion across them.

The simulation employs a timestep $\Delta t = 0.1 \tau_A$ (which corresponds to $\Delta t \chi_\parallel = 2$) and a uniform logical grid resolution of $128\times64\times32$, partitioned into 512 parallel domains. The simulation is initialized with a small applied radial velocity perturbation, $V_r(r,\theta,t=0) = 10^{-4}\sin{(\pi r)}\cos{(\theta)}$, where $r$ here is the logical radial coordinate and $\theta$ the logical poloidal-angle coordinate.
The number of GMRES iterations needed per timestep to achieve a relative residual of $10^{-3}$ is 25.5, averaged over the entire simulation duration of 16,277 timesteps. The iteration and
$\Delta t/\Delta t_{exp}$ histories are provided in Fig. \ref{fig:cfl-history}-right, with similar conclusions as with the Bennett test.

\begin{figure}
    \centering
    \includegraphics[scale=0.17]{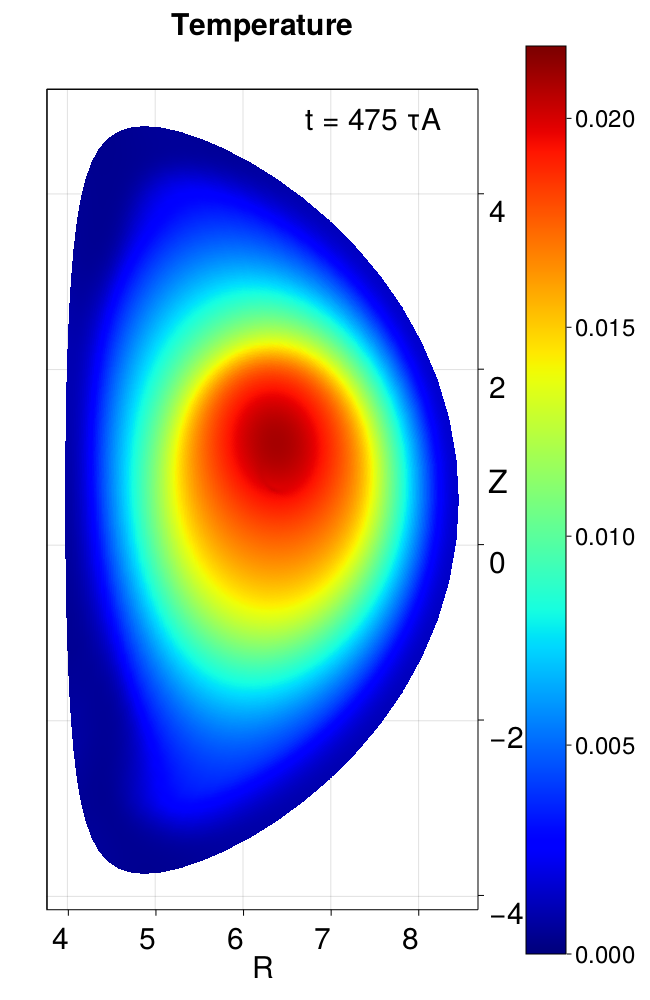}
    ~\includegraphics[scale=0.17]{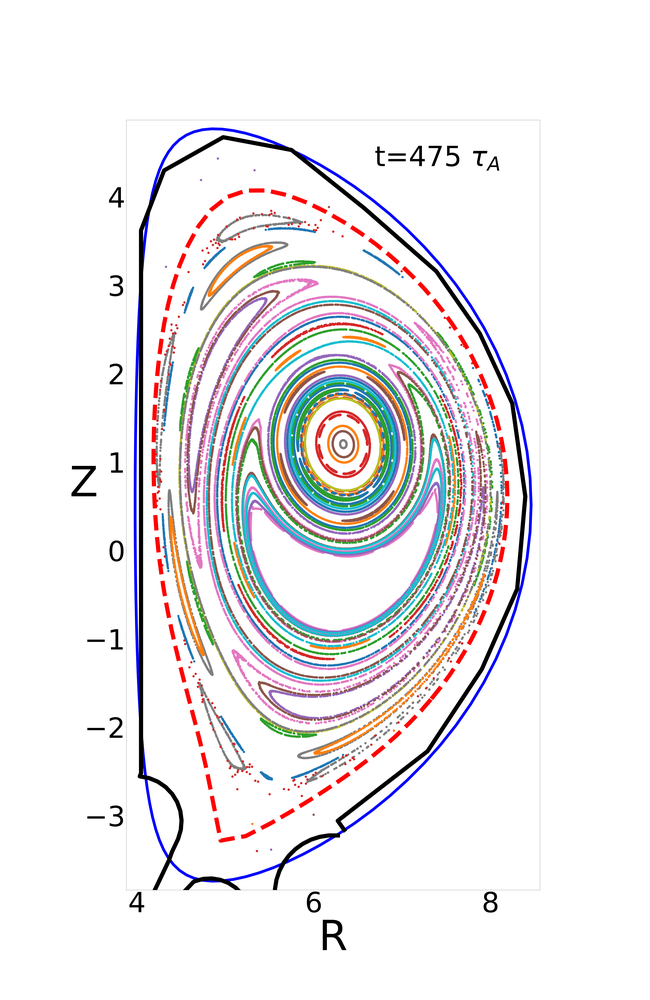}
    ~\includegraphics[scale=0.17]{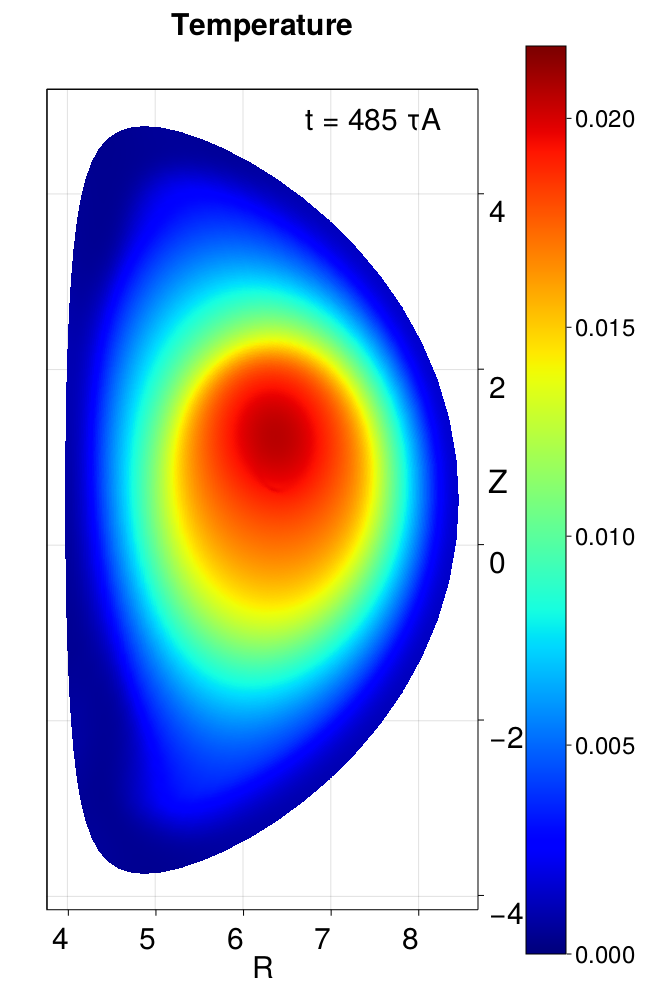}
    ~\includegraphics[scale=0.17]{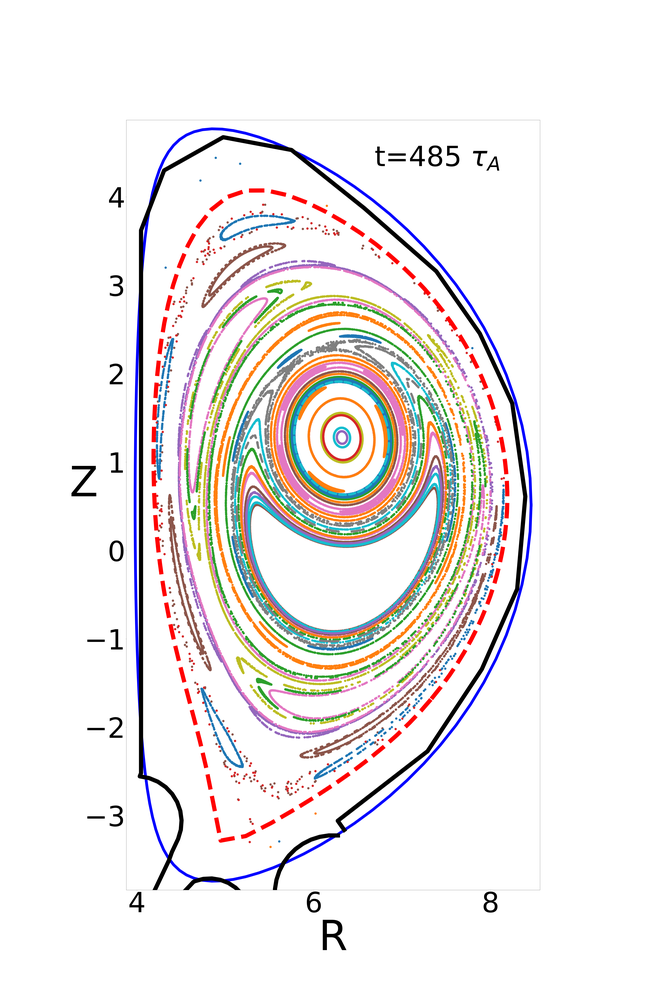}\\
    \includegraphics[scale=0.17]{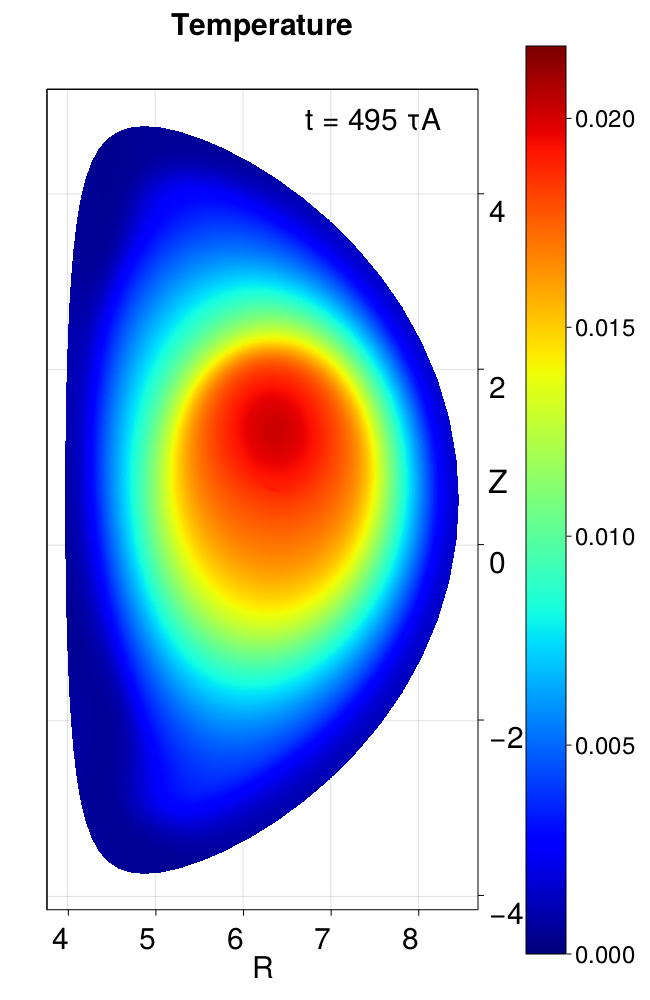}
    ~\includegraphics[scale=0.17]{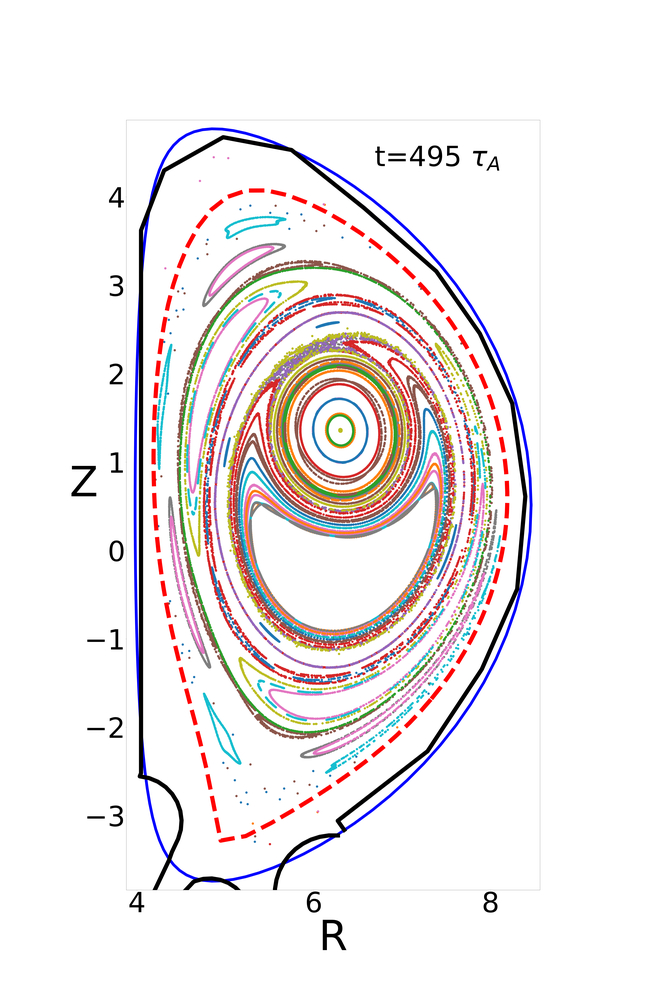}
    ~\includegraphics[scale=0.17]{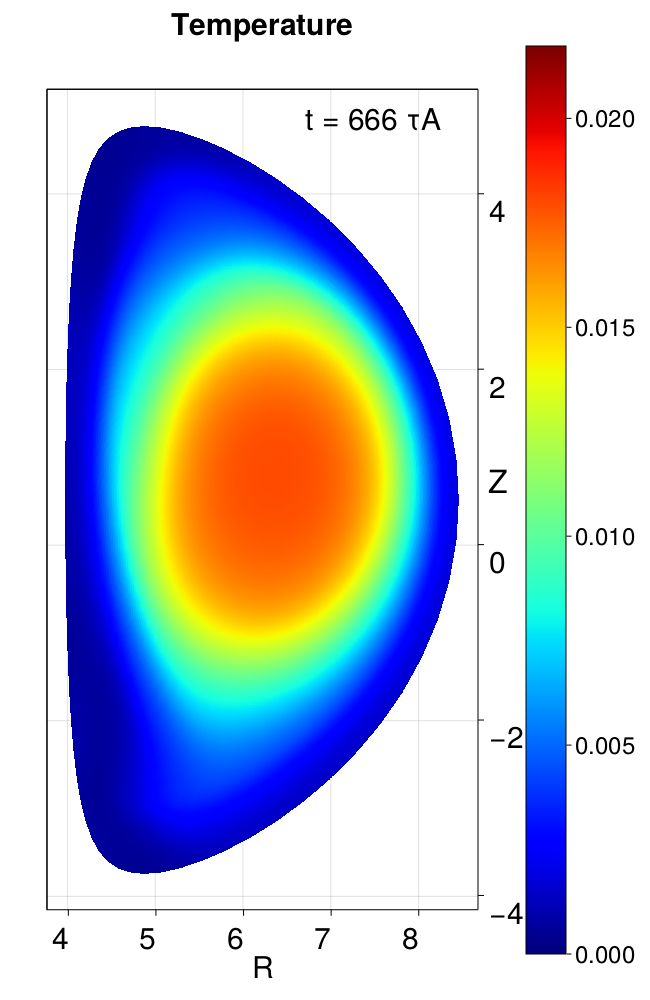}
    ~\includegraphics[scale=0.17]{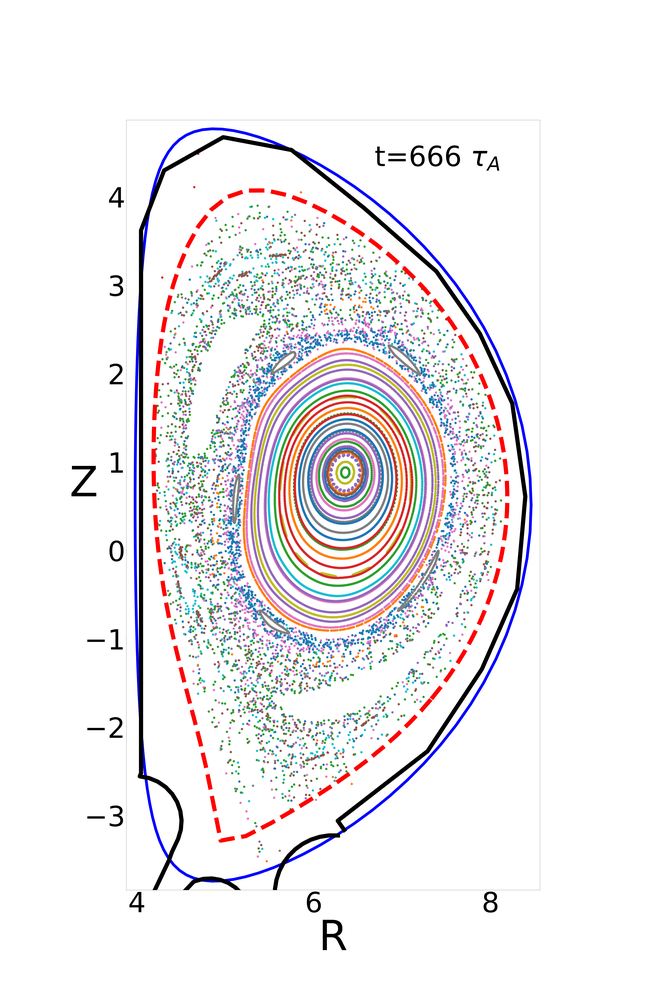}
    \caption{ITER test: Snapshots of the temperature  (shown with the color map) and magnetic topology (shown with the Poincar\'e maps) before, during, and after the reconnection event, with a resolution of 128x64x32. Timestamps are provided in the figures. The temperature remains confined to the closed flux surfaces.}
    \label{fig:ITER11}
\end{figure}

The evolving magnetic topology and temperature field during the ensuing disruption are shown via Poincare sections and temperature plots in Figure \ref{fig:ITER11}, respectively.
As the (1,1) kink mode develops, the thermal energy of the core is seen to remain within the region of closed flux surfaces, as expected. This result highlights the suitability of our method for realistic 3D MHD applications in realistic geometries and highly dynamic scenarios.

\section{Conclusions}

\label{sec:Discussion-and-conclusions}We have demonstrated a practical
high-order scheme for the strongly anisotropic heat transport equation
that is easy to implement in existing finite-difference implementations,
is fourth-order accurate, is robust, and offers manageable numerical
errors for realistically large anisotropies (with up to $10^{7}$
considered in this study). Critically, the approach is amenable to
modern multigrid solvers for implicit time integration, rendering
the implementation scalable under grid refinement (i.e., with the
number of GMRES iterations per timestep very weakly dependent on the
number of mesh points and transport anisotropy), and scaling as $\sqrt{\Delta t\chi_{\parallel}}$
with the implicit timestep. The approach has been demonstrated with
well-accepted targeted numerical tests, as well as integrated MHD
simulations of kink instabilities in magnetically confined tokamak fusion
reactors in both idealized (helical) and realistic (toroidal) geometries.
We conclude that the method offers a viable, practical and efficient
numerical scheme to deal with the strongly anisotropic transport equation
in realistic plasma conditions and geometries.

\section*{Acknowledgments}

The authors acknowledge useful conversations with X. Tang and J. N. Shadid. This work was supported by Triad National Security, LLC under contract
89233218CNA000001, and the DOE
Office of Applied Scientific Computing Research (ASCR) through the DOE FES-ASCR SciDAC Partnership Program and the ASCR Applied Mathematics Program. The research
used computing resources provided by the Los Alamos National Laboratory
Institutional Computing Program.

\appendix

\section{Implementation details of the fourth-order accurate discretization}

\label{app:discrete-details}

The full discrete treatment of the anisotropic tensor-diffusion operator
for both inner-domain and boundary points is as follows. We begin
with a conservative divergence discretization in 2D (with straightforward
generalization to 3D):
\begin{eqnarray*}
\nabla\cdot\left(\bar{\bar{\Xi}}\cdot\nabla\varphi\right)|_{i,j} & = & \frac{1}{\mathit{\Delta x}}\left[\left(\Xi^{xx}\frac{\partial\varphi}{\partial x}+\Xi^{xy}\frac{\partial\varphi}{\partial y}\right)_{i+1/2,j}-\left(\Xi^{xx}\frac{\partial\varphi}{\partial x}+\Xi^{xy}\frac{\partial\varphi}{\partial y}\right)_{i-1/2,j}\right]\\
 & + & \frac{1}{\mathit{\Delta y}}\left[\left(\Xi^{yy}\frac{\partial\varphi}{\partial y}+\Xi^{xy}\frac{\partial\varphi}{\partial x}\right)_{i,j+1/2}-\left(\Xi^{yy}\frac{\partial\varphi}{\partial y}+\Xi^{xy}\frac{\partial\varphi}{\partial x}\right)_{i,j-1/2}\right].
\end{eqnarray*}
For the co-derivative fluxes at faces, we interpolate the face-centered
flux from cell-centered values of $\varphi$ to fourth-order accuracy
as follows:

{\small
\begin{eqnarray*}
\left(\Xi^{xx}\frac{\partial\varphi}{\partial x}\right)_{i+1/2,j} & = & \frac{1}{\Delta x}\begin{cases}
C_{1\times5}^{+}\Xi_{N_{x}-3:N_{x}+1,j}^{xx}A_{2:6,6}\varphi_{N_{x}-4:N_{x}+1,j}, & i=N_{x}\\
C_{1\times4}^{0}\Xi_{N_{x}-2:N_{x}+1,j}^{xx}A_{3:6,6}\varphi_{N_{x}-4:N_{x}+1,j}, & i=N_{x}-1\\
C_{1\times4}^{0}\Xi_{i-1:i+2,j}^{xx}A_{2:5,6}\varphi_{i-2:i+3,j}, & 2\leq i\leq N_{x}-2\\
C_{1\times4}^{0}\Xi_{0:3,j}^{xx}A_{1:4,6}\varphi_{0:5,j}, & i=1\\
C_{1\times5}^{-}\Xi_{0:4,j}^{xx}A_{1:5,6}\varphi_{0:5,j}, & i=0
\end{cases},
\end{eqnarray*}
}and similarly for $\left(\Xi^{yy}\frac{\partial\varphi}{\partial y}\right)_{i,j+1/2}$,
with:
\begin{align*}
C_{1\times4}^{0}=\frac{1}{12}\left(\begin{matrix}-1, & 7, & 7, & -1\end{matrix}\right) & ,\\
C_{1\times5}^{+}=\frac{1}{12}\left(\begin{matrix}2, & 17, & -11, & 5, & -1\end{matrix}\right) & ,\\
C_{1\times5}^{-}=\frac{1}{12}\left(\begin{matrix}-1, & 5, & -11, & 17, & 2\end{matrix}\right) & ,
\end{align*}
and:
\[
A_{6\times6}=\frac{1}{60}\left(\begin{matrix}-137 & 300 & -300 & 200 & -75 & 12\\
-12 & -65 & 120 & -60 & 20 & -3\\
3 & -30 & -20 & 60 & -15 & 2\\
-2 & 15 & -60 & 20 & 30 & -3\\
3 & -20 & 60 & -120 & 65 & 12\\
-12 & 75 & -200 & 300 & -300 & 137
\end{matrix}\right).
\]
For the cross-derivative fluxes, a similar fourth-order interpolation
yields:{\small
\begin{eqnarray*}
\left(\Xi^{xy}\frac{\partial\varphi}{\partial y}\right)_{i+1/2,j} & = & \begin{cases}
C_{1\times5}^{+}\left(\Xi^{xy}\frac{\partial\varphi}{\partial y}\right)_{N_{x}-3:N_{x}+1,j}, & i=N_{x}\\
C_{1\times4}^{0}\left(\Xi^{xy}\frac{\partial\varphi}{\partial y}\right)_{i-1:i+2,j}, & 1\leq i<N_{x}-1\\
C_{1\times5}^{-}\left(\Xi^{xy}\frac{\partial\varphi}{\partial y}\right)_{0:4,j}, & i=0
\end{cases},
\end{eqnarray*}
with:
\[
\left(\Xi^{xy}\frac{\partial\varphi}{\partial y}\right)_{i,j}=\frac{\Xi_{i,j}^{xy}}{\mathit{\Delta y}}\begin{cases}
D_{1\times5}^{0}\varphi_{i,j-2:j+2}^{T}, & 2\leq j\leq N_{y}-1\\
D_{1\times5}^{-}\varphi_{i,0:5}^{T}, & j=0,1\\
D_{1\times5}^{+}\varphi_{i,N_{y}-3:N_{y}+1}^{T}, & j=N_{y},N_{y}+1
\end{cases},
\]
and:
\begin{align*}
D_{1\times5}^{0}=\frac{1}{12}(1,-8,0,8,-1) & ,\\
D_{1\times5}^{+}=\frac{1}{12}(-1,6,-18,10,3) & ,\\
D_{1\times5}^{-}=\frac{1}{12}(-3,-10,18,-6,1) & .
\end{align*}
}{\small\par}

\section{Logical-to-physical coordinate map in the ITER simulation}

\label{app:iter-mesh}

The ITER kink mode simulation of Section \ref{sec:iter} uses a logical coordinate system $\left(r,\theta,\phi\right)\,\in\,\left[0,1\right]\times\left[0,2\pi\right)\times\left[0,2\pi\right)$.
The logical coordinates are mapped to cylindrical coordinates using the transformation:
\begin{align*}
R\left(r,\theta,\phi\right)=& R_m + (R_0 -R_m) r + a r \cos{\left[\theta+\arcsin{\left(\delta \, r^2 \sin{\theta}\right)}\right]} ,\\
Z\left(r,\theta,\phi\right)=& Z_m + (Z_0 - Z_m) r + a \left(r\kappa + (1-r)\kappa_s\right) r \sin{\left[\theta+\zeta \,r^2 \sin\left(2\theta\right)\right]},\\
\phi_c\left(\phi\right)=& -\phi.
\end{align*}
The shaping parameters for the ITER experiment are: minor radius $a=2.24$m, geometric axis $R_0=6.219577546$m, $Z_0=0.5143555944$m, magnetic axis $R_m=6.341952203$m, $Z_m=0.6327986088$m, triangularity $\delta=0.6$, elongation $\kappa=1.9$, elongation at the magnetic axis $\kappa_s=1.35$, and squareness $\zeta=0.06$.
We emphasize that the cylindrical angle $\phi_c$ rotates about the origin in the opposite direction as the toroidal angle $\phi$.


\bibliographystyle{ieeetr}
\bibliography{ref,numerics,transport,numerical_MHD,general}

\end{document}